\documentclass[english]{article}
\usepackage[T1]{fontenc}
\usepackage[latin9]{inputenc}
\usepackage{amsmath}
\usepackage{amssymb}
\usepackage{graphicx}
\usepackage{esint}
\usepackage{color}

\makeatletter
\newcommand{\lyxaddress}[1]{
\par {\raggedright #1
\vspace{1.4em}
\noindent\par}
}

\usepackage{a4wide}
\usepackage{babel}

\newcommand{\On}[1]{\mathrm{O}(#1)}
\newcommand{\on}[1]{\mathfrak{o}(#1)}
\newcommand{\gn}{\mathfrak{g}}
\newcommand{\hn}{\mathfrak{h}}
\newcommand{\fn}{\mathfrak{f}}
\newcommand{\n}{\mathbf{n}}
\newcommand{\nt}{\tilde{\mathbf{n}}}
\newcommand{\nh}{\hat{\mathbf{n}}}
\newcommand{\dd}{\mathrm{d}}

\newcommand{\Ac}{\mathrm{A}}
\newcommand{\Kc}{\mathrm{K}}
\newcommand{\Ad}[1]{\mathrm{A}^{(#1)}}
\newcommand{\Kd}[1]{\mathrm{K}^{(#1)}}
\newcommand{\As}[1]{A^{(#1)}}
\newcommand{\Ks}[1]{K^{(#1)}}
\newcommand{\Lc}{\mathrm{L}}
\newcommand{\tr}{\mathrm{tr}}
\newcommand{\ac}{\mathrm{a}}

\newcommand{\SU}[1]{\mathrm{SU}(#1)}
\newcommand{\Un}[1]{\mathrm{U}(#1)}

\newcommand{\SO}[1]{\mathrm{SO}(#1)}

\newcommand{\ui}[2]{u^{(#1)}_{#2}}

\newcommand{\vxi}{\vec{\xi}}
\newcommand{\vr}{\vec{r}}
\newcommand{\vs}{\vec{s}}

\makeatother

\usepackage{babel}
\begin{document}

\title{On integrable boundaries in the 2 dimensional $O(N)$ $\sigma$-models}

\author{Ines Aniceto$^{1}$, Zoltán Bajnok$^{2}$, Tamás Gombor$^{2,3}$,
Minkyoo Kim$^{2}$\\
 and László Palla$^{3}$}

\maketitle

\lyxaddress{\begin{center}
$^{1}$ \emph{Institute of Physics, Jagiellonian University, }\\
\emph{Ul. \L ojasiewicza 11, 30-348 Kraków, Poland }\\
\emph{$^{2}$MTA Lend\"ulet Holographic QFT Group, Wigner Research
Centre for Physics,}\\
\emph{Konkoly-Thege Miklós u. 29-33, 1121 Budapest , Hungary}\\
$^{3}$\emph{Institute for Theoretical Physics, Roland Eötvös University,
}\\
\emph{ 1117 Budapest, Pázmány Péter sétány 1/A, Hungary}
\par\end{center}}
\begin{abstract}
We make an attempt to map the integrable boundary conditions for 2
dimensional non-linear O(N) $\sigma$-models. We do it at various
levels: classically, by demanding the existence of infinitely many
conserved local charges and also by constructing the double row transfer
matrix from the Lax connection, which leads to the spectral curve
formulation of the problem; at the quantum level, we describe the
solutions of the boundary Yang-Baxter equation and derive the Bethe-Yang
equations. We then show how to connect the thermodynamic limit of
the boundary Bethe-Yang equations to the spectral curve. 
\end{abstract}

\section{Introduction}

Integrable quantum field theories are useful toy examples of particle
physics. Their popularity is due to the fact that many physical quantities
can be calculated exactly and, despite their simplicity, they exhibit
phenomena relevant for QCD. In particular, 2 dimensional (2D) $O(N)$
$\sigma$-models are asymptotically free in perturbation theory and
their classical conformal invariance is broken by a dynamically generated
mass scale $\Lambda$. Massive excitations form the vector multiplet
of the $O(N)$ group with factorized scattering \cite{Zamolodchikov:1978xm},
which makes possible to calculate the relation between the mass $m$
and the parameter $\Lambda$ \cite{Hasenfratz:1990ab}. 

The $O(N)$ $\sigma$-models are also relevant from the AdS/CFT point
of view. In a large class of integrable string $\sigma$-models strings
propagate on the product of an anti-de Sitter space and spheres $S^{n}$
\cite{Beisert:2010jr}. Light-cone gauge fixed string theories on
the sphere part are described classically by the $O(N)$ $\sigma$-models.
In the string theory applications we are often interested in open
strings, strings ending on some D-brane submanifolds of $S^{N}$ \cite{Zoubos:2010kh}.
This translates to $O(N)$ models with boundaries, and an important
question is to classify those boundary conditions which maintain integrability.
This is the motivation of our work. 

Interestingly, there are not many papers analyzing integrable boundary
conditions for $O(N)$ models. Soon after the seminal paper of Ghoshal
and Zamolodchikov on integrable boundaries \cite{Ghoshal:1993tm}
Ghoshal determined the solutions of the boundary Yang-Baxter equation
(BYBE) in the $O(N)$ $\sigma$-models with diagonal reflections having
$O(N)$ and $O(N-1)$ symmetries \cite{Ghoshal:1994bc}. These reflection
factors correspond to free and fixed boundary conditions for the fundamental
fields. Later Corrigan and Sheng established the classical integrability
of the free boundary condition by constructing infinitely many conserved
charges via the Lax connection \cite{Corrigan:1996nt}. They also
found a new (field dependent) boundary condition in the $O(3)$ model.
Using the boundary generalizations of the Goldschmidt-Witten argument,
Moriconi and de Martino \cite{Moriconi:1998gc} indicated that free
and fixed boundary conditions can be quantum integrable and even extended
the result for a mixture of free and fixed boundary conditions (for
the boundary value of the fundamental field). Later Moriconi analyzed
systematically the boundary conditions of the $O(N)$ models \cite{Moriconi:2001mc,Moriconi:2001xz}.
He identified new types of integrable boundary conditions, which can
be implemented by adding a quadratic boundary potential including
the time derivative of the fundamental field to the Lagrangian. He
managed to transform the $O(3)$ boundary condition found by Corrigan
and Sheng to his class. These boundary conditions can be represented
by an antisymmetric matrix, which can be brought into a 2 by 2 block-diagonal
form. Thus they break the $O(N)$ symmetry to the products of $O(2)$s.
He then searched for the quantum analogues of this new class of boundary
conditions and found a few non-diagonal representatives only. Namely
only with a single block and Dirichlet boundary conditions, or in
$O(2N)$ models with all 2 by 2 block being the same. This classification
was confirmed and extended in the $O(4)$ case to a two parameter
family of reflection factors in \cite{Arnaudon:2003gj}. It was argued
in \cite{He:2003iw} that the non-diagonal boundary conditions does
not have a consistent Hamiltonian description, although, as we will
show the constraint was not properly implemented in the Lagrangian
description. The open boundary integrability in the string theory,
relevant for the $O(N)$ models, was analyzed in \cite{Mann:2006rh,Dekel:2011ja,Bajnok:2013sza}.
It seemed from the investigations that not all combinations of free
and fixed boundary conditions are compatible with Lax integrability.
Similar conclusion was drawn by investigating integrable 
boundary conditions in coset theories \cite{MacKay:2001bh,MacKay:2004rz}.
The aim of our paper is to investigate the integrability of boundary
conditions at various levels: Lagrangian, Lax, quantum, trying to
map them as completely as possible and to establish their relations. 

The rest of this paper is organized as follows: In section \ref{sec:Classical-integrability}
we analyze the integrability at the classical level. We start by reviewing
the construction of conserved charges in the periodic case. We use
three different descriptions: in the first we use the stereographically
projected coordinates on $S^{N-1}$, in the second we use the embedding
coordinates in $\mathbb{R}^{N}\supset S^{N-1}$, while in the third
we consider the sphere $S^{N-1}$ as a coset $\frac{SO(N)}{SO(N-1)}$
theory. In the second part of section \ref{sec:Classical-integrability}
we use these descriptions to map the integrable boundary conditions
in the model, while also showing the equivalence of the various descriptions.
At the end of this section we formulate the boundary integrability
at the language of the Lax connection. We construct the double row
transfer matrix and use its eigenvalues to define the spectral curve.
We analyze the analytical and symmetry properties of this curve, which
provides an alternative way to find and classify classical solutions
and also helps in quantizing the model. We close the section by explicitly
constructing the spectral curve of some rotating string solutions
in the $O(4)$ model. Section \ref{sec:Quantum-integrability} is
devoted to the quantum theory. Integrable boundary conditions at the
quantum level are charachterized by reflection matrices, which solve
the boundary Yang-Baxter equations and satisfy unitarity and boundary
crossing unitarity. We systematically describe these reflection matrices
and use them to derive the Bethe-Yang equations, which provide the
asymptotic spectrum on a large interval. At the end of the section
we calculate the classical limit of the spectrum for two boundary
conditions in the $O(4)$ model and reproduce the classical spectral
curve. We conclude in section \ref{sec:Conclusions}, while technical
details are relegated to two appendices.

\section{Classical integrability\label{sec:Classical-integrability}}

In this section we investigate the classical integrability of the
$O(N)$ $\sigma$-model in the presence of boundaries. We start by
introducing three different descriptions in the bulk theory, then
following the same descriptions for the theory with boundaries.

\subsection{Bulk formulations}

Here we recall the bulk formulations using unconstrained and constrained
variables on the sphere. We finish by regarding the sphere as a coset.

\subsubsection{Unconstrained fields}

The $O(N)$ $\sigma$-model is a 2D field theory of a variable which
lives on an $N-1$-dimensional sphere. This sphere is naturally given
by the unit sphere in $\mathbb{R}^{N}$ and can be projected stereographically
onto the hyperplane passing through the origin. Let us denote the
corresponding projected coordinates by $\vxi=(\xi^{1},\dots,\xi^{N-1})$.
The Lagrangian of the model is given by 
\begin{equation}
\mathcal{L}=4\frac{\partial_{\alpha}\vxi\cdot\partial^{\alpha}\vxi}{(1+\xi^{2})^{2}}\qquad;\qquad\xi^{2}=\vxi\cdot\vxi\,,\label{eq:Lxi}
\end{equation}
with the metric being the pullback of the flat metric in $\mathbb{R}^{N}$.
Variation of the action gives the equations of motion:
\begin{equation}
\partial_{\mu}\partial^{\mu}\xi^{j}-\frac{4\,\partial^{\mu}\xi^{j}\,\vxi\cdot\partial_{\mu}\vxi-2\xi^{j}\,\partial_{\mu}\vxi\cdot\partial^{\mu}\vxi}{1+\xi^{2}}=0.\label{eq:eomxi}
\end{equation}
The model has an explicit $O(N-1)$ symmetry, as the currents 
\begin{equation}
J_{\alpha}^{ij}=\frac{4}{(1+\xi^{2})^{2}}(\xi^{i}\partial_{\alpha}\xi^{j}-\xi^{j}\partial_{\alpha}\xi^{i})\label{eq:aram1}
\end{equation}
are conserved on shell. In fact, the model has an implicit $O(N)$
symmetry. The extra conserved symmetry currents are 
\begin{equation}
J_{\alpha}^{iN}=-J_{\alpha}^{Ni}=-\frac{2}{(1+\xi^{2})^{2}}(2\xi^{i}\vxi\cdot\partial_{\alpha}\vxi+(1-\xi^{2})\partial_{\alpha}\xi^{i}),\label{eq:aram2}
\end{equation}
and correspond to the infinitesimal transformations 
\[
\xi^{j}=\xi^{j}+\epsilon(2\xi^{j}\xi^{i}+(1-\xi^{2})\delta^{ij}),\quad j=1,\dots,N-1\,.
\]
These transformations change $\xi^{2}$ by $\delta\xi^{2}=2\epsilon\xi^{i}(1+\xi^{2})$. 

The energy-momentum tensor of the model is traceless, and its lightcone
components are 
\[
T_{++}=\frac{4}{(1+\xi^{2})^{2}}\sum\limits _{i}(\partial_{+}\xi^{i})^{2}=-\sum\limits _{I,J=1}^{N}J_{+}^{IJ}J_{+}^{JI}\,,
\]
and similarly for $T_{--}$. Here we introduced light-cone coordinates
$\sigma_{\pm}\equiv\frac{1}{2}\left(\tau\pm\sigma\right)$ and $\partial_{\pm}\equiv\partial_{\tau}\pm\partial_{\sigma}$.
Conservation of the currents implies that 
\[
\partial_{-}T_{++}=0\to T_{++}(\sigma^{+})\quad;\qquad\partial_{+}T_{--}=0\to T_{--}(\sigma^{-}).
\]
Thus, $\sigma$-models are classically conformal and possess infinitely
many conservation laws: 
\begin{equation}
\partial_{-}T_{++}^{k}=0\quad;\qquad\partial_{+}T_{--}^{k}=0\quad;\qquad k>2.\label{eq:cl_cons_law}
\end{equation}
Indeed, the conserved currents associated to the spin $\pm s$ charges
$Q_{\pm s}$ must satisfy

\begin{equation}
\partial_{-}T^{(s+1)}=\partial_{+}\Theta^{(s-1)}\quad;\qquad\partial_{-}\bar{T}^{(s+1)}=\partial_{+}\bar{\Theta}^{(s-1)}\label{eq:Bulk_cons}
\end{equation}
Thus choosing $T^{(2s)}=T_{++}^{k}$, $\Theta^{(2s-2)}=0$ and $\bar{T}^{(2s)}=T_{--}^{k}$,
$\bar{\Theta}^{(2s-2)}=0$ leads to conserved higher spin charges
\begin{equation}
Q_{2k-1}=\int T_{++}^{k}(\sigma,\tau)d\sigma\quad;\qquad Q_{-2k+1}=\int T_{--}^{k}(\sigma,\tau)d\sigma\quad;\qquad k>1
\end{equation}
which are algebraically independent of the stress tensor (whose charges
are given by $k=1$). 
In the $O(N)$ $\sigma$-models there are higher polynomial expressions of the currents leading to higher 
spin conserved charges \cite{Evans:2000qx}. They are related to Casimirs of the $O(N)$ group and form a Poisson commuting 
set. In this paper we do not study the Poisson structure and focus only on the charges related to the conformality of
the model, i.e. to $T_{\pm\pm}^k$.

\subsubsection{Constrained fields}

To make the full $O(N)$ symmetry manifest we can use coordinates
on $\mathbb{R}^{N}:$
\begin{equation}
n^{i}=\frac{2\xi^{i}}{1+\xi^{2}},\quad i=1,\dots,N-1,\qquad n^{N}=\frac{1-\xi^{2}}{1+\xi^{2}}\,.\label{eq:nxirel}
\end{equation}
They parametrize the unit sphere as $\n^{t}=(n^{1},\dots,n^{N})\in\mathbb{R}^{N}$
with the constraint $\n^{t}\n=1$. This constraint has to be included
in the Lagrangian to maintain equivalence with eq. (\ref{eq:Lxi}):
\begin{align*}
\mathcal{L} & =\partial_{a}\n^{t}\partial^{a}\n-\lambda(\n^{t}\n-1)\,.
\end{align*}
Variation of the action leads to the equation of motion 
\[
\partial_{a}\partial^{a}\n+\lambda\n=0.
\]
Using the constraint, the Lagrange multiplier can be eliminated leading
to 
\[
\partial_{a}\partial^{a}\n+(\partial_{a}\n^{t}\partial^{a}\n)\n=0\,,
\]
which is equivalent to eq. (\ref{eq:eomxi}) once the relation (\ref{eq:nxirel})
is used. The conserved currents take the universal form 
\[
J_{\alpha}^{IJ}=n^{I}\partial_{\alpha}n^{J}-n^{J}\partial_{\alpha}n^{I}\,,
\]
and the energy momentum tensor is given by 
\[
T_{\pm\pm}=\partial_{\pm}\n\cdot\partial_{\pm}\n\,.
\]
From the unit vector $\n$ one can define the group element
\[
m=\mathbb{I}-2\n\n^{t}
\]
with $m^{-1}=m$, such that the current one-form reads as
\begin{equation}
J=m\,\dd m=-2(\dd\n)\n^{t}+2\n\,\dd\n^{t}.\label{eq:Jm}
\end{equation}
In terms of this current one-form, the Lagrangian is simply
\[
\mathcal{L}=\tr(J_{\alpha}J^{\alpha})=\tr(J\wedge\star J)\,.
\]
The flatness of the current, together with its conservation 
\[
dJ+J\wedge J=0\quad;\qquad d\star J=0\,,
\]
can be packed into the flatness of a spectral parameter-dependent
Lax connection:
\begin{equation}
\ac(\lambda):=\frac{1}{1-\lambda^{2}}J-\frac{\lambda}{1-\lambda^{2}}\star J\,.\label{eq:Jconnection}
\end{equation}
These formulas together with (\ref{eq:Jm}) resemble the formulation
of a coset theory. Indeed, $S^{N-1}$ can be represented as an $\On{N}/\On{N-1}$
coset, which leads to the following description.

\subsubsection{Gauged $\sigma$-model point of view}

A map between the $\On{N}/\On{N-1}:=\{g\sim gh|g\in\On{N},h\in\On{N-1}\}$ coset
and the sphere $S^{N-1}:=\{\n\in\mathbb{R}^{N}|\n^{t}\n=1\}$ can
be obtained by choosing a representative point $\n_{0}^{t}=\{1,0,\dots,0\}$
on $S^{N-1}$. The $\On{N-1}$ subgroup, which leaves $\n_{0}$ invariant
is the $\left(N-1\right)\times\left(N-1\right)$ lower right corner
of $O(N)$, whose Lie algebra is denoted by $\hn$. The map between
the coset and the sphere is simply 
\begin{align*}
\frac{\On{N}}{\On{N-1}} & \rightarrow S^{N-1}\qquad;\qquad gh\rightarrow g\n_{0}\,.
\end{align*}
The Maurer-Cartan form, 
\[
\omega=g^{-1}\dd g=\Ac+\Kc,
\]
can be decomposed w.r.t. the coset structure as $\omega\in\on{N}=\hn\oplus\fn$
with $\Ac\in\hn,\Kc\in\fn$, where $\Kc$ contains the physical degrees
of freedom and $\Ac$ is a gauge field. By definition, this current
satisfies the flatness condition: 
\[
\dd\omega+\omega\wedge\omega=0.
\]
Using the properties 
\[
[\hn,\hn]=\hn\quad;\quad[\hn,\fn]=\fn\quad;\qquad[\fn,\fn]=\hn\,,
\]
the flatness condition for $\omega$ can be decomposed as 
\begin{align}
 & \dd\Ac+\Ac\wedge\Ac+\Kc\wedge\Kc=0\,;\label{eq:flat1}\\
 & \dd\Kc+\Ac\wedge\Kc+\Kc\wedge\Ac=0\,.\label{eq:flat2}
\end{align}
One can introduce the operators which project onto the $\hn$ and
$\fn$ subspaces as follows: 
\begin{align*}
\Pi_{\hn}:\on{N} & \rightarrow\hn\qquad;\qquad v\rightarrow\frac{1}{2}(v+jvj)\,,\\
\Pi_{\fn}:\on{N} & \rightarrow\fn\qquad;\qquad v\rightarrow\frac{1}{2}(v-jvj)\,,
\end{align*}
 where $j=\mathbb{I}-2\n_{0}\n_{0}^{t}$. The gauge invariant Lagrangian
takes the form: 
\begin{equation}
\mathcal{L}=\tr\left[\Pi_{\fn}(\omega_{a})\Pi_{\fn}(\omega^{a})\right]=\tr(K_{a}K^{a}).\label{eq:SK}
\end{equation}
 To obtain the equations of motion one can make the variations
\begin{align*}
g & \rightarrow g(1+\epsilon)\quad;\qquad\omega\rightarrow\omega+\dd\epsilon+[\omega,\epsilon].
\end{align*}
 where $\epsilon\in\fn$ since $\epsilon\in\hn$ would not change
the action. This variation changes the action by 
\[
\delta\mathcal{L}=2\tr[(\partial_{a}\epsilon+[A_{a},\epsilon])K^{a}]=-2\tr[\epsilon(\partial_{a}K^{a}+[A_{a},K^{a}])]+2\partial_{a}\tr[\epsilon K^{a}]\,,
\]
and leads to the equation of motion 
\begin{align}
\dd\star\Kc+\Ac\wedge\star\Kc+\star\Kc\wedge\Ac=0\,.\label{eq:EOM}
\end{align}
To make contact with the formulation of the constrained field $\n$
we recall that 
\begin{align}
\n & =g\n_{0}\quad;\qquad m=gjg^{t}\quad;\qquad J=g(j\omega j-\omega)g^{t}=-2g\Pi_{\fn}(\omega)g^{t}=-2g\Kc g^{t}\in g\fn g^{t}.\label{eq:coset-constr}
\end{align}
This makes the two formulations completely equivalent. Finally, we
note that the equations of motion can be encoded into the flatness
of a spectral parameter-dependent Lax connection:

\begin{align}
\Lc(\lambda)=\Ac+\frac{\lambda^{2}+1}{\lambda^{2}-1}\Kc-\frac{2\lambda}{\lambda^{2}-1}\star\Kc.\label{eq:LaxGauge}
\end{align}

\subsection{Boundary formulations}

Let us turn to the formulations of the boundary problem in the same
order as they were analyzed in the bulk theory.

\subsubsection{Unconstrained fields}

Using the unconstrained fields, the boundary theory can obtained by
restricting the space coordinates to an open interval parametrized
by $\sigma\in(0,\pi)$ 
\begin{equation}
S=\int d\tau\int\limits _{0}^{\pi}d\sigma\frac{\partial_{\alpha}\vxi\cdot\partial^{\alpha}\vxi}{(1+\xi^{2})^{2}}.\label{eq:Sxibdry}
\end{equation}
Now, because of the boundaries, when computing the variations of the
action $\xi^{k}\rightarrow\xi^{k}+\delta\xi^{k}$, $\xi^{i}\rightarrow\xi^{i}$
($i\neq k$), the following surface terms arise 
\begin{equation}
-\frac{2\delta\xi^{k}\partial_{\sigma}\xi^{k}}{(1+\xi^{2})^{2}}\vert_{0}^{\pi}.\label{eq:surface}
\end{equation}
If there is no constraint at the boundary for $\xi^{k}$ then there
is no summation over $k$. (If there were any constraints, they should
be added to the Lagrangian with a Lagrange multiplier). Assuming that
there is no long-range interaction between the boundaries, the surface
terms must vanish separately, i.e. we find the consistent boundary
conditions (b.c.-s) 
\begin{equation}
\delta\xi^{k}\partial_{\sigma}\xi^{k}=0,\qquad k=1,\dots,N-1\label{eq:bcxi}
\end{equation}
on both ends of the interval. If we interpret the conditions $\delta\xi^{k}\vert_{0}=0$
(or $\delta\xi^{k}\vert_{\pi}=0$) as also implying the vanishing
of $\partial_{\tau}\xi^{k}$, then we conclude that the consistent
b.c.-s imply either Neumann or Dirichlet b.c.-s for the fields $\xi^{i}$.
Let us focus on the integrability of these b.c.-s. According to \cite{Ghoshal:1993tm},
the bulk conservation laws (\ref{eq:Bulk_cons}) lead to a boundary
conserved quantity if, at the boundary, the difference 
\begin{equation}
\left[T^{(s+1)}-\bar{T}^{(s+1)}+\bar{\Theta}^{(s-1)}-\Theta^{(s-1)}\right]\vert=\frac{d\Sigma}{dt}
\end{equation}
is a total time-derivative of some quantity $\Sigma$. 
(By the
empty vertical line we mean to evaluate the expression at the boundary.
When the boundary is not specified, then the statement is true for
both boundaries.)  In the most
general case $\Sigma$ can even depend on the dynamical fields. In
this paper we restrict our analysis only to the cases when $\Sigma=0$.
Clearly, the boundary conditions (\ref{eq:bcxi}) we found  are conformal 
and guarantee the vanishing of
$(T_{++}^{k}-T_{--}^{k})\vert\propto(T_{++}-T_{--})\vert\sim\frac{\partial_{\tau}\vxi\cdot\partial_{\sigma}\vxi}{(1+\xi^{2})^{2}}\vert=0$
for all $k.$ This ensures the existence of infinitely many conserved
higher spin charges, which are independent of the energy. 
Still it may happen that the infinite number of conserved charges following
from the conformality of the boundary condition is not "infinite enough" to
ensure integrability. In particular conserved charges should form a commuting family. 
It was shown in \cite{MacKay:2004rz} that coset boundary conditions compatible with 
the bulk coset structure lead to infinite number of conserved charges in involution.  Since in
the framework of this paper we do not investigate either the Poisson structure or the 
higher Casimir charges we have no tool to check if conformality implies integrability or not. 

Let us try to extend the Lagrangian with a boundary potential, which
could preserve integrability of the model. Motivated by previous investigations
\cite{Moriconi:2001xz}, we add also a boundary Lagrangian term $\sum\limits _{I,J}n^{I}M_{IJ}\dot{n}^{J}$
with an antisymmetric matrix $M_{IJ}=-M_{JI}$ (and where we used
$\dot{\ }\equiv\partial_{\tau}{\ }$). Here capital indices $I,J$
run from $1$ to $N$, while lower case indices $i,j$ run from 1
to $N-1$. One readily obtains 
\[
\sum\limits _{I,J}n^{I}M_{IJ}\dot{n}^{J}=-\frac{1}{2}{\rm tr}(MJ_{\tau})=\frac{2}{(1+\xi^{2})^{2}}\biggl(\sum\limits _{ij}M_{ij}(\xi^{i}\dot{\xi}^{j}-\xi^{j}\dot{\xi}^{j})-\sum\limits _{i}\bigl(2\xi^{i}\vxi\cdot\dot{\vxi}+(1-\xi^{2})\dot{\xi}^{i}\bigr)M_{iN}\biggr)\ .
\]
We compute the change of this boundary piece under the $\xi^{k}\rightarrow\xi^{k}+\delta\xi^{k}$
variation, with the understanding that in every term containing $\dot{\delta\xi^{k}}$
we integrate by parts and drop the integrated terms. This way one
finds 
\begin{eqnarray*}
\delta{\rm tr}(MJ_{\tau}) & = & \frac{16\delta\xi^{k}}{(1+\xi^{2})^{2}}\Bigl(-\frac{\xi^{k}}{1+\xi^{2}}\bigl(\sum\limits _{ij}M_{ij}(\xi^{j}\dot{\xi}^{i}-\xi^{i}\dot{\xi}^{j})+2\sum\limits _{i}M_{iN}\dot{\xi}^{i}\bigr)\,+\\
 &  & \qquad\,\,\,\,\,\hspace*{1em}\,\,\,\,\,\sum\limits _{i}M_{ik}\dot{\xi}^{i}+\frac{2\vxi\cdot\dot{\vxi}}{1+\xi^{2}}(M_{kN}+\sum\limits _{j}M_{kj}\xi^{j})\Bigr)\,.
\end{eqnarray*}
Combining this with the surface terms (\ref{eq:surface}) coming from
the variation of the bulk action, one finds the boundary conditions
($k=1,\dots,N-1$) 
\begin{eqnarray}
\partial_{\sigma}\xi^{k}\vert & = & -\Bigl(-\frac{\xi^{k}}{1+\xi^{2}}\bigl(\sum\limits _{ij}M_{ij}(\xi^{j}\dot{\xi}^{i}-\xi^{i}\dot{\xi}^{j})+2\sum\limits _{i}M_{iN}\dot{\xi}^{i}\bigr)+\sum\limits _{i}M_{ik}\dot{\xi}^{i}+\label{eq:teljes}\\
 &  & \,\,\,\,\,\,\,\,\,\,\,\,\,\,\,\,\,\,\,\,\,\,\frac{2\vxi\cdot\dot{\vxi}}{1+\xi^{2}}(M_{kN}+\sum\limits _{j}M_{kj}\xi^{j})\Bigr)\biggr\rvert\,.\nonumber 
\end{eqnarray}
This boundary condition is conformal, as direct calculation guarantees
that $\partial_{\tau}\vxi\cdot\partial_{\sigma}\vxi\propto(T_{++}^{k}-T_{--}^{k})\vert=0$,
providing infinitely many higher spin charges. Let us analyse the same
boundary conditions in the alternative formulations.

\subsubsection{Constrained fields}

The action which corresponds to the theory (\ref{eq:Sxibdry}) in
terms of the constrained variable $\n$ reads as

\begin{equation}
S=\int d\tau\int\limits _{0}^{\pi}d\sigma\left[\partial_{a}\n^{t}\partial^{a}\n-\lambda(\n^{t}\n-1)+\delta(\sigma)\lambda_{0}(\n^{t}\n-1)-\delta(\sigma-\pi)\lambda_{\pi}(\n^{t}\n-1)\right]\,.\label{eq:Snbdry}
\end{equation}
Observe that we have implemented the constraint $\n^{t}\n$ also at
the boundary.\emph{ All previous analysis seemed to miss this term}.
Variation now leads to the bulk equation of motion and to the boundary
condition: 
\[
\delta\n^{t}(\partial_{\sigma}\n-\lambda_{\vert}\n)\vert=0\,.
\]
Thus for any $i$ we can choose either generalized Neumann or Dirichlet
boundary conditions. We shall assume that $l$ directions satisfy
generalized Neumann boundary condition while $N-l$ directions obey
Dirichlet instead. 
\[
\partial_{\sigma}n_{i}\vert_{0}=\lambda_{0}n_{i}\vert_{0}\quad;\qquad i=1,\dots,l\quad;\qquad\delta n_{i}\vert_{0}=0\quad;\qquad i=l+1,\dots,N\,.
\]
All these boundary conditions are conformal 
\[
(T_{++}-T_{--})\vert\sim\partial_{\tau}\n^{t}\cdot\partial_{\sigma}\n\vert=0.
\]
and conformality also guarantees that
\begin{equation}
(T_{++}^{k}-T_{--}^{k})\vert\propto(T_{++}-T_{--})\vert=0\,,
\end{equation}
thus infinitely many higher spin conserved charges exist. 

Without loss of generality, we can choose the Dirichlet directions
as 
\[
n_{l+1}\vert_{0}=\alpha\quad;\qquad n_{l+2}\vert_{0}=\dots=n_{N}\vert_{0}=0\,.
\]
This implies that $\sum_{i=1}^{l}n_{i}n_{i}\vert_{0}=1-\alpha^{2}$,
which can be used to determine $\lambda_{0}$ and obtain the boundary
condition for the generalized Neumann directions:
\begin{equation}
\partial_{\sigma}n_{i}\vert_{0}=\frac{n_{i}}{1-\alpha^{2}}\sum_{j=1}^{l}n_{j}\partial_{\sigma}n_{j}\vert_{0}\,.\label{eq:gen_neum_bc}
\end{equation}
These boundary conditions are equivalent to Dirichlet and Neumann
boundary conditions for appropriately rotated $\xi_{i}$ variables,
see Appendix \ref{sec:Symm-boundary-conds} for the details.

A special case is when $\alpha=0$, i.e. we restrict the boundary
field to a sphere of maximal radius. This can be obtained by intersecting
the unit sphere with a hyperplane passing through the origin. This
boundary condition is given by the $\xi_{i}=0$ Dirichlet boundary
conditions. Actually, in this case the space-derivative of the constraint
$\n^{t}\n=1$ implies that 
\[
\sum_{i=1}^{l}n_{i}\partial_{\sigma}n_{i}\bigr\rvert_{0}=0\quad\longrightarrow\qquad\partial_{\sigma}n_{i}\vert_{0}=0\quad;\quad i=1,\dots,l\,,
\]
and thus the remaining directions satisfy Neumann boundary condition.
We analyze the symmetry of this boundary condition in detail in Appendix
\ref{sec:Symm-boundary-conds}. It turns out that the symmetry is
$O(l)\times O(N-l)$. 

In order to get the most general conformal boundary condition, we
could demand that the time and space derivatives of $\n$ are orthogonal
at the boundary: $\partial_{\tau}\n^{t}\cdot\partial_{\sigma}\n\vert=0$.
This can be achieved by adding a boundary potential with an antisymmetric
matrix $M$ to the boundary Lagrangian: 
\[
\mathcal{L}_{b}=\delta(\sigma)\left(\n^{t}M\partial_{\tau}\n+\lambda_{0}(\n^{t}\n-1)\right),
\]
where, for definiteness, we added it at the $\sigma=0$ boundary.
Similar terms could be added at $\sigma=\pi$ as well. Here we again
emphasize that the constraint had to be added to the boundary piece,
as required by consistency. After eliminating the Lagrange multiplier,
the boundary condition turns out to be
\begin{equation}
M\partial_{\tau}\n-(\n^{t}M\partial_{\tau}\n)\n=\partial_{\sigma}\n.\label{eq:BCMn}
\end{equation}
Contracting with $\n^{t}$ on the left we can see that this is indeed
consistent with the constraint $\n^{t}\n=1$, in contrast to what
one can find in the literature, where the b.c. appears without the
second term \cite{Moriconi:2001xz,He:2003iw}. Using eq. (\ref{eq:nxirel}),
it is straightforward to show that the boundary condition in eq. (\ref{eq:teljes}),
given in terms of the unconstrained variables, is equivalent to this
one in terms of the $\n$ fields. Conformality of
the boundary conditions follow from $\partial_{\tau}\n^{t}\cdot\partial_{\sigma}\n\vert=0$.
It is also instructive to rewrite the boundary condition for the current
and group elements. In terms of the current, the boundary term reads
simply as
\[
\mathcal{L}_{b}=\delta(\sigma)\tr\left(J_{\tau}M\right)\,.
\]
Making an infinitesimal variation
\[
m\rightarrow m(1+\epsilon),\quad;\qquad J\to J+d\epsilon+[J,\epsilon]\,,
\]
with $\epsilon$ satisfying the constraint (i.e. it is an element
of the $g\fn g^{t}$ subspace), changes the bulk part of the action
as 
\[
\delta\mathcal{L}=\tr[(\partial_{a}\epsilon+[J_{a},\epsilon])J^{a}]=-\tr[\epsilon(\partial_{a}J^{a})]+\partial_{a}\tr[\epsilon J^{a}]\,,
\]
while the boundary part changes as 
\[
\delta\mathcal{L}_{b}=\tr\left\{ (\partial_{\tau}\epsilon+[J_{\tau},\epsilon])M\right\} =\tr(\epsilon[M,J_{\tau}])=\tr(\epsilon[g\Pi_{\hn}(g^{t}Mg)g^{t},J_{\tau}]).
\]
 Therefore, the boundary condition is 
\[
J_{\sigma}=[g\Pi_{\hn}(g^{t}Mg)g^{t},J_{\tau}]=\frac{1}{2}([M,J_{\tau}]+[mMm,J_{\tau}]).
\]
 This particularly nice boundary condition is explicitly conformal, as 
\[
(T_{++}^{k}-T_{--}^{k})\vert\sim(T_{++}-T_{--})\vert\sim\tr(J_{\sigma}J_{\tau})
\]
We show in appendix \ref{sec:Symm-boundary-conds} that this is equivalent
to the boundary condition we got from the $\xi$ variables. 

This whole analysis can be also recovered from the gauge theory formulation,
which follows.

\subsubsection{Gauged $\sigma$-model point of view}

Boundary conditions in the coset language are geometrical. In particular,
in the case of integrable boundary conditions of principal chiral
models, the group element is restricted to a coset \cite{MacKay:2001bh,MacKay:2004rz}.
Thus, we first analyze the boundary conditions by restricting the
boundary field to $S^{N-1}\rightarrow S^{N-k-1}$, where the spheres
have radius 1 and there is no extra boundary Lagrangian term in the
coset action. In the language of the $\n$ this means 
\[
n_{i}=0,\quad\text{where}\ i=N-k+1,\dots,N.
\]
 We introduce new notation by decomposing $\n$ as: $\n=\nt+\nh$,
with 
\[
\nt=(n_{1},\dots,n_{N-k},0,\dots,0)\quad,\qquad\nh=(0,\dots,0,n_{N-k+1},\dots,n_{N}),
\]
The boundary conditions then become:
\[
\partial_{1}\nt=0\quad;\qquad\partial_{0}\nh=0\,.
\]
 Let us derive this boundary condition using the coset language. At
the boundary, the only non-zero variables are the $\nt$-s. Introduce
an $\On{N-k}\times\On{k}$ subgroup of $\On{N}$, which can be used
for the parametrization of the $S^{N-k-1}$ subspace. We will denote
it by $G_{1}$, and its Lie algebra by $\gn_{1}$. Let us also denote
the little subgroup of $G_{1}$ by $H_{1}$ and its Lie algebra by
$\hn_{1}$. The fields at the boundary can be parametrized by $S^{N-k-1}\equiv G_{1}/H_{1}:=\{g_{1}\sim g_{1}h_{1}|g_{1}\in G_{1},h_{1}\in\ H_{1}\}$
and $\nt=g_{1}\n_{0}$ where $g_{1}\in G_{1}$. We will have to use
a decomposition of $\hn$, $\fn$ and $\gn_{1}$: 
\begin{align*}
 & \hn=\hn_{1}\oplus\hn_{2}, &  & \fn=\fn_{1}\oplus\fn_{2}, &  & \hn_{1}\oplus\fn_{1}=\gn_{1};\\
 & [\hn_{1},\hn_{1}]\subset\hn_{1}, &  & [\hn_{1},\fn_{1}]\subset\fn_{1}, &  & [\fn_{1},\fn_{1}]\subset\hn_{1}, & [\hn_{1},\hn_{2}]\subset\hn_{2}, &  & [\hn_{1},\fn_{2}]\subset\fn_{2},\\
 & [\hn_{2},\hn_{2}]\subset\hn_{1}, &  & [\hn_{2},\fn_{1}]\subset\fn_{2}, &  & [\hn_{2},\fn_{2}]\subset\fn_{1}, & [\fn_{1},\fn_{2}]\subset\hn_{2}, &  & [\fn_{2},\fn_{2}]\subset\hn_{1}.
\end{align*}
 The $\on{N}=\gn$ algebra w.r.t. to the splitting above becomes:
\[
\gn=\begin{pmatrix}0 & \fn_{1} & \fn_{2}\\
\fn_{1} & \hn_{1} & \hn_{2}\\
\fn_{2} & \hn_{2} & \hn_{1}
\end{pmatrix}.
\]
 This is a $\mathbb{Z}_{2}\times\mathbb{Z}_{2}$ graded algebra with
$\{\hn_{1},\hn_{2},\fn_{1},\fn_{2}\}\sim\{(0,0),(0,1),(1,0),(1,1)\}$.
Thus, the two symmetric cosets $G/H$ and $G/G_{1}$ are compatible,
where $G=\On{N}$.

We can observe that the physical currents live in the $\fn_{1}\sim(1,0)$
subspace which is the even part of the $G/H$ and the odd part of
the $G/G_{1}$ decompositions. Therefore ''coset of the bulk\textquotedbl{}
and \textquotedbl{}coset of the boundary'' mean different things.

The decomposition of the current at the boundary is 
\[
\omega=g^{t}\dd g=\Ad{1}+\Ad{2}+\Kd{1}+\Kd{2},
\]
 where $\Ad{i}\in\hn_{1}$ and $\Kd{i}\in\fn_{1}$.

At the boundary, $g\in G_{1}$, so $\As{2}_{\tau}=\Ks{2}_{\tau}=0$.
This is equivalent to the boundary conditions 
\[
[A_{\tau},\kappa]=0\quad;\qquad[K_{\tau},\kappa]=0,
\]
 where 
\[
\kappa=\mathrm{diag}(1,\dots,1,-1,\dots,-1).
\]
 When we make the variation of the action (\ref{eq:SK}), we have
to use $g\rightarrow g(1+\epsilon)$ with $\epsilon\in\gn_{1}$ at
the boundary. After the variation we get the $\Ks{1}_{\sigma}=0$
boundary conditions, which are equivalent to 
\begin{align*}
\{K_{\sigma},\kappa\} & =0.
\end{align*}
Let us now add a gauge-invariant boundary term to the Lagrangian,
of the form 
\[
\mathcal{L}_{b}=\tr[\Pi_{\fn}(\omega_{\tau})\Pi_{\fn}(g^{t}Mg)],
\]
 where $M\in\on{n}$ is a constant matrix. After the variation we
get: 
\[
\delta\mathcal{L}_{b}=\tr\left\{ (\partial_{\tau}\epsilon+[A_{\tau},\epsilon])\Pi_{\fn}(g^{t}Mg)+K_{\tau}[\Pi_{\hn}(g^{t}Mg),\epsilon]\right\} =2\tr(\epsilon[K_{\tau},\Pi_{\hn}(g^{t}Mg)])+\partial_{\tau}\tr(\epsilon\Pi_{\fn}(g^{t}Mg)).
\]
 Using also terms from the variation of the bulk part of the action,
we arrive at the boundary condition: 
\[
K_{\sigma}=[\Pi_{\hn}(g^{t}Mg),K_{\tau}].
\]
This boundary condition is conformal as 
\[
(T_{++}-T_{--})\vert\propto\tr(K_{\sigma}K_{\tau})=0\,.
\]
In terms of components we have
\[
\As{2}_{0}=0\quad;\qquad\Ks{2}_{0}=0\quad;\qquad\Ks{1}_{1}=\Pi_{\gn_{1}}\left([\Pi_{\hn}(g^{t}Mg),K_{0}]\right).
\]
This result is consistent with the result obtained in the language
of the constrained fields $\n$: 
\begin{align*}
\partial_{\tau}\nh=0\quad;\qquad\partial_{\sigma}\nt & =M\partial_{\tau}\nt-(\nt^{t}M\partial_{\tau}\nt)\nt.
\end{align*}

\subsection{Lax connection and spectral curve}

In this section we construct Lax matrices for the boundary problem,
which lead to its spectral curve formulation.

\subsubsection{Bulk transfer matrix}

We have already mentioned two different versions of the Lax connections.
In the formulation based on the fundamental field it was defined by
\[
\ac(\lambda):=\frac{1}{1-\lambda^{2}}J-\frac{\lambda}{1-\lambda^{2}}\star J\,,
\]
while in the gauged $\sigma$-model formulation we found 
\[
\Lc(\lambda)=\Ac+\frac{\lambda^{2}+1}{\lambda^{2}-1}\Kc-\frac{2\lambda}{\lambda^{2}-1}\star\Kc,
\]
where $\omega=g^{t}\dd g=\Ac+\Kc$. By recalling the relation 
\begin{align*}
J & =g(j\omega j-\omega)g^{t}=-2g\Pi_{\fn}(\omega)g^{t}=-2g\Kc g^{t}\,,
\end{align*}
 we can see that the two connections are related by a ``gauge''
transformation:
\[
\Lc(\lambda)=\Ac+\frac{\lambda^{2}+1}{\lambda^{2}-1}\Kc-\frac{2\lambda}{\lambda^{2}-1}\star\Kc=\omega-2(\frac{1}{1-\lambda^{2}}\Kc-\frac{\lambda}{1-\lambda^{2}}\star\Kc)=g^{t}\dd g+g^{t}\ac(\lambda)g\,.
\]
Consequently, gauge-invariant quantities can be easily expressed in
any of these formulations. The usefulness of the Lax connection lies
in the fact that one can generate from it an infinite family of conserved
charges. One first defines the transport matrix 
\begin{equation}
T\left(b,a,\lambda\right)=\mathcal{\mathfrak{\mathcal{P}}}\overleftarrow{\exp}\biggl\{-\int_{a}^{b}\mathrm{d}\sigma\,\ac_{\sigma}\left(\sigma,\lambda\right)\biggr\},\label{eq:Bulk-monodromy-def}
\end{equation}
for a path connecting $a$ and $b$. The transport matrix of the other
connection $\mathbf{L}(\lambda)$, can then be expressed as 
\begin{equation}
\mathcal{\mathfrak{\mathcal{P}}}\overleftarrow{\exp}\biggl\{-\int_{a}^{b}\mathrm{d}\sigma\,\mathbf{L}_{\sigma}\left(\sigma,\lambda\right)\biggr\}=g^{t}(b)T(b,a,\lambda)g(a)\label{eq:gaugetransfT}
\end{equation}

In the cylindrical geometry (bulk theory) one can integrate the connection
for a spacial non-contractible loop, such as for the path from $0$
to $2\pi$, leading to a gauge-invariant quantity, $T(2\pi,0,\lambda)$,
called the monodromy matrix. Using the flatness condition of the connection
$\ac_{a}$: $\partial_{\tau}\boldsymbol{\ac}_{\sigma}=\partial_{\sigma}\boldsymbol{\ac}_{\tau}-\left[\boldsymbol{\ac}_{\sigma},\boldsymbol{\ac}_{\tau}\right]$,
we can calculate the time derivative of the transport matrix 
\begin{eqnarray}
\partial_{\tau}T\left(b,a,\lambda\right) & = & -\int_{a}^{b}\mathrm{d\sigma}\partial_{\sigma}\biggl(\mathcal{\mathfrak{\mathcal{P}}}\overleftarrow{\exp}\Bigl\{-\int_{\sigma}^{b}\mathrm{d}\sigma^{\prime}\,\ac_{\sigma}\left(\sigma^{\prime},\lambda\right)\Bigr\}\ac_{\tau}\left(\sigma,\lambda\right)\mathcal{\mathfrak{\mathcal{P}}}\overleftarrow{\exp}\Bigl\{-\int_{a}^{\sigma}\mathrm{d}\sigma^{\prime\prime}\,\ac_{\sigma}\left(\sigma^{\prime\prime},\lambda\right)\Bigr\}\biggr)\nonumber \\
 & = & T\left(b,a,\lambda\right)\ac_{\tau}\left(a,\lambda\right)-\ac_{\tau}\left(b,\lambda\right)T\left(b,a,\lambda\right).\label{eq:dtTransport}
\end{eqnarray}
 Since $\ac_{\tau}(2\pi,\lambda)=\ac_{\tau}(0,\lambda)$ this ensures
that the trace of the monodromy matrix, called the transfer matrix,
is time-independent 
\[
\mathbf{T}\left(\lambda\right)=\tr\mathcal{\mathfrak{\mathcal{P}}}\overleftarrow{\exp}\biggl\{-\int_{0}^{2\pi}\mathrm{d}\sigma\,\ac_{\sigma}\left(\sigma,\lambda\right)\biggr\}\,,
\]
and generates infinitely many conserved charges. The bulk theory was
thoroughly analysed in \cite{Beisert:2004ag}, in the context of AdS/CFT
(see also \cite{Beisert:2005bm} for the supersymmetric equivalent).
Let us now turn to the parallel construction in the presence of boundaries.

\subsubsection{Boundary transfer matrix}

The monodromy matrix in the boundary case takes a double row type
form \cite{Mann:2006rh,Dekel:2011ja}
\[
\Omega\left(\lambda\right)=U_{0}\left(\lambda\right)T_{R}\left(2\pi,\pi,\lambda\right)U_{\pi}\left(\lambda\right)T\left(\pi,0,\lambda\right),
\]
where $U_{0,\pi}(\lambda)$ are (as of yet $\lambda$ and time-dependent)
$O(N)$ matrices encoding the type of boundary conditions we have
at $\sigma=0,\pi$, respectively, and the matrix $T_{R}\left(2\pi,\pi,\lambda\right)$
is the reflected transport matrix, obtained via a parity transformation
$\sigma\rightarrow2\pi-\sigma$. Taking into account that parity transforms
the currents as%
\footnote{The minus sign in the reflected current $J_{\sigma}^{R}\left(\sigma\right)$
comes from the fact that $J_{\sigma}$ includes a derivative which
under parity transforms as $\partial_{\sigma}\rightarrow-\partial_{\sigma}$.%
}
\[
J_{\sigma}\left(\sigma\right)\rightarrow J_{\sigma}^{R}\left(\sigma\right)=-J_{\sigma}\left(2\pi-\sigma\right)\quad;\qquad J_{\tau}\left(\sigma\right)\rightarrow J_{\tau}^{R}\left(\sigma\right)=J_{\tau}\left(2\pi-\sigma\right),
\]
we can see that
\[
\ac_{\sigma}^{R}\left(\sigma,\lambda\right)=-\ac_{\sigma}\left(2\pi-\sigma,-\lambda\right).
\]
The reflected transport matrix is then given by
\begin{eqnarray}
T_{R}\left(2\pi,\pi,\lambda\right) & \equiv & \mathcal{\mathfrak{\mathcal{P}}}\overleftarrow{\exp}\left\{ -\int_{\pi}^{2\pi}\mathrm{d}\sigma\,\ac_{\sigma}^{R}\left(\sigma,\lambda\right)\right\} =\mathcal{\mathfrak{\mathcal{P}}}\overleftarrow{\exp}\left\{ -\int_{\pi}^{0}\mathrm{d}\sigma\,\ac_{\sigma}\left(\sigma,-\lambda\right)\right\} \label{eq:reflected-bulk-monodromy-def}\\
 & = & T\left(\pi,0,-\lambda\right)^{-1}.\nonumber 
\end{eqnarray}
This leads to the following form of the monodromy matrix 
\begin{equation}
\Omega\left(\lambda\right)=U_{0}\left(\lambda\right)T\left(-\lambda\right)^{-1}U_{\pi}\left(\lambda\right)T(\lambda)\quad;\qquad T(\lambda)=T\left(\pi,0,\lambda\right).\label{eq:Monodromy-def}
\end{equation}
The most general condition of integrability can be written in terms
of $\Omega(\lambda)$ as 
\begin{equation}
\partial_{\tau}\Omega(\lambda)=[N(\lambda),\Omega(\lambda)]
\end{equation}
with some appropriate $N(\lambda),$ as this condition guarantees
that ${\rm tr}(\Omega(\lambda)^{k})$ is conserved for any integer
$k$. Thus, expanding the boundary transfer matrix $T(\lambda)={\rm tr(\Omega(\lambda))}$
in the spectral parameter generates infinitely many conserved charges. 

Using eq. (\ref{eq:dtTransport}) we can calculate the time derivative
of the boundary monodromy matrix as 
\begin{eqnarray*}
\partial_{\tau}\Omega\left(\lambda\right) & = & \partial_{\tau}U_{0}(\lambda)U_{0}\left(\lambda\right)^{-1}\Omega\left(\lambda\right)-U_{0}\left(\lambda\right)\ac_{\tau}\left(0,-\lambda\right)U_{0}\left(\lambda\right)^{-1}\Omega\left(\lambda\right)+\Omega\left(\lambda\right)\ac_{\tau}\left(0,\lambda\right)+\\
 &  & +U_{0}\left(\lambda\right)T\left(-\lambda\right)^{-1}\bigl(\partial_{\tau}U_{\pi}(\lambda)-U_{\pi}\left(\lambda\right)\ac_{\tau}\left(\pi,\lambda\right)+\ac_{\tau}\left(\pi,-\lambda\right)U_{\pi}\left(\lambda\right)\bigr)T\left(\lambda\right).
\end{eqnarray*}
Demanding that 
\begin{eqnarray}
\partial_{\tau}U_{0}(\lambda) & = & U_{0}\left(\lambda\right)\ac_{\tau}\left(0,-\lambda\right)-\ac_{\tau}\left(0,\lambda\right)U_{0}\left(\lambda\right);\nonumber \\
\partial_{\tau}U_{\pi}(\lambda) & = & U_{\pi}\left(\pi\right)\ac_{\tau}\left(\pi,\lambda\right)-\ac_{\tau}\left(\pi,-\lambda\right)U_{\pi}\left(\lambda\right),\label{eq:Ui-restrictions-time-evolution}
\end{eqnarray}
the time evolution of the monodromy matrix is given by 
\begin{equation}
\partial_{\tau}\Omega\left(\lambda\right)=\left[\Omega\left(\lambda\right),\ac_{\tau}\left(0,\lambda\right)\right].
\end{equation}
Thus integrability requires the existence of matrices $U_{0}$ and
$U_{\pi}$ which satisfy eq. (\ref{eq:Ui-restrictions-time-evolution}).
These matrices are not gauge invariant. Indeed, using eq. (\ref{eq:gaugetransfT})
we can see that the boundary matrix $U_{i}$ in the $\mathbf{L}$
formulation is $\tilde{U}_{i}=gU_{i}g^{t}$ such that
\begin{align}
[A_{0},\tilde{U}] & =0\qquad;\qquad[K_{0},\tilde{U}]=0\qquad;\qquad\{K_{1},\tilde{U}\}=0.\label{eqA}
\end{align}
 This actually shows that constant $U$ matrices in one description
lead to time-dependent matrices in the other. Classification of all
$U$ matrices satisfying eq. (\ref{eq:Ui-restrictions-time-evolution})
means the classification of the integrable boundary conditions in
the Lax language. Let us start with the investigations of time-independent
$U$-s.In terms of the currents, the time-independent restrictions
become 
\begin{eqnarray*}
\left.U_{0}\left(\lambda\right)\left(J_{\tau}-\lambda J_{\sigma}\right)-\left(J_{\tau}+\lambda J_{\sigma}\right)U_{0}\left(\lambda\right)\right|_{\sigma=0} & = & 0;\\
\left.U_{\pi}\left(\lambda\right)\left(J_{\tau}+\lambda J_{\sigma}\right)-\left(J_{\tau}-\lambda J_{\sigma}\right)U_{\pi}\left(\lambda\right)\right|_{\sigma=\pi} & = & 0.
\end{eqnarray*}
Assuming a Taylor expansion for the matrices $U_{i}\left(\lambda\right)\in O(N)$:
\begin{equation}
U_{i}\left(\lambda\right)=\sum_{n=0}^{+\infty}U_{i}^{(n)}\lambda^{n},\label{eq:Ui-Taylor-exp}
\end{equation}
 we can solve the above restrictions order by order in powers of $\lambda$.
We easily obtain 
\begin{eqnarray}
\left.\left[U_{i}^{(0)},J_{\tau}\right]\right|_{\sigma=i} & = & 0,\; i=0,\pi;\\
\left.\left[U_{0}^{(k)},J_{\tau}\right]\right|_{\sigma=0} & = & \left.\left\{ U_{0}^{(k-1)},J_{\sigma}\right\} \right|_{\sigma=0}\quad;\qquad\left.\left[U_{\pi}^{(k)},J_{\tau}\right]\right|_{\sigma=\pi}=-\left.\left\{ U_{\pi}^{(k-1)},J_{\sigma}\right\} \right|_{\sigma=\pi},\; k\ge1.\nonumber 
\end{eqnarray}
In the principal chiral model we found $\lambda$ dependent $U-s$,
but in the $\mathrm{{\rm O}}(N)$ $\sigma$-model we could manage
to find only constant ones. In this case, the requirements become
\begin{equation}
\left.\left[U_{i},J_{\tau}\right]\right|_{\sigma=i}=\left.\left\{ U_{i},J_{\sigma}\right\} \right|_{\sigma=i},\quad i=0,\pi.\label{eq:Ui-commutation-Js}
\end{equation}
Using the definition of the currents $J_{\pm}$, it is easy to check
that at the boundary we have 
\begin{equation}
\left.J_{\pm}\right|_{\sigma=i}=U_{i}^{-1}\,\left.J_{\mp}\right|_{\sigma=i}U_{i},\: i=0,\pi.\label{eq:Ui-commutation-jpm}
\end{equation}
and that $\left.\left[J_{\pm},U_{i}^{2}\right]\right|_{\sigma=i}=0,\; i=0,\pi.$
Naturally, this condition is satisfied with the condition $U_{i}^{2}=\boldsymbol{1}$.
This actually means that $J_{\pm}\to\alpha(J_{\pm})=UJ_{\pm}U^{-1}$
defines an automorphism of the Lie algebra leaving the energy momentum
tensor invariant: $T_{++}=\tr(J_{+}^{2})=T_{--}=\tr(J_{-}^{2})$.
Automorphisms, which satisfy eq. (\ref{eq:Ui-commutation-Js}), are
well-known for Lie algebras and are related to their symmetric space
decompositions \cite{MR1834454}: 
\[
\mathfrak{o}(N)=\mathfrak{h}+\mathfrak{m}\quad;\qquad\alpha(\mathfrak{h})=\mathfrak{h}\quad;\qquad\alpha(\mathfrak{m})=-\mathfrak{m}\,,
\]
such that 
\[
[\mathfrak{h},\mathfrak{h}]\subset\mathfrak{h}\quad,\qquad[\mathfrak{h},\mathfrak{m}]\subset\mathfrak{m}\quad;\qquad[\mathfrak{m},\mathfrak{m}]\subset\mathfrak{h}\,.
\]
The various components of the currents at the boundary have to be
in different subspaces
\[
J_{\tau}\vert\in\mathfrak{h}\quad;\qquad J_{\sigma}\rvert\in\mathfrak{m}\,,
\]
 Choosing the decomposition 
\[
\mathfrak{h}=\mathfrak{so}(k)\oplus\mathfrak{so}(N-k)\quad;\qquad\mathfrak{m}=\frac{\mathfrak{so}(N)}{\mathfrak{so}(N)\oplus\mathfrak{so}(N-k)}
\]
leads to the matrix 
\begin{equation}
U={\rm diag}(\underbrace{1,\dots,1}_{k},\underbrace{-1,\dots,-1}_{N-k})\label{eq:U-Neumann-Dirichlet}
\end{equation}
The corresponding boundary condition is exactly what we can describe
in all cases as 
\[
\partial_{\sigma}n_{i}\vert=0\quad;\qquad i=1,\dots,k\qquad;\qquad n_{i}\vert=0\quad;\qquad i=k+1,\dots,N\,,
\]
 i.e. Neumann for the first $k$ components and vanishing Dirichlet
for the remaining $N-k$. This also means that the group element $m$
is restricted to the coset $\frac{O(N)}{O(k)\times O(N-k)}$ and $[U,m]=0$. 

In the following we analyze the analytic properties of the eigenvalues
of the monodromy matrix for this case (\ref{eq:U-Neumann-Dirichlet}).

\subsection{Analytic properties of the boundary transfer matrix}

In the following we assume that $U_{0}=U_{\pi}=U$ is a $\lambda$
independent constant such that $U^{2}=1$. By definition, the transport
matrix is an element of the ${\rm O}(N)$ group. Since an open path
can be contracted to a point, it has to be in the identity component
of ${\rm O}(N)$, i.e. in $SO(N)$. Taking a generic complex $\lambda$,
the transport matrix sits in $SO(N,\mathbb{C})$, and so will also
be the the monodromy matrix. Then, by diagonalization we can bring
it into the form 
\begin{equation}
\Omega_{\mathrm{diag}}\left(\lambda\right)=\begin{cases}
\mathrm{diag}\left(\mathrm{e}^{iq_{1}\left(\lambda\right)},\mathrm{e}^{-iq_{1}\left(\lambda\right)},\cdots,\mathrm{e}^{iq_{[N/2]}\left(\lambda\right)},\mathrm{e}^{-iq_{[N/2]}\left(\lambda\right)}\right) & ,\;\mbox{for}\: N\,\mbox{even}\\
\mathrm{diag}\left(\mathrm{e}^{iq_{1}\left(\lambda\right)},\mathrm{e}^{-iq_{1}\left(\lambda\right)},\cdots,\mathrm{e}^{iq_{[N/2]}\left(\lambda\right)},\mathrm{e}^{-iq_{[N/2]}\left(\lambda\right)},1\right) & ,\;\mbox{for}\: N\,\mbox{odd}
\end{cases}.\label{eq:quasimomenta-SON}
\end{equation}
The eigenvalues come in pairs, and the $q_{i}\left(\lambda\right)$
are the so-called quasi-momenta. They parametrize a multi-sheeted
Riemann surface, which has a specific structure for each classical
solution. Common features for all are the pole singularities at $\lambda=\pm1$
and branch cuts, starting whenever two quasi-momenta coincide. Let
us denote by $\mathcal{C}_{r}$ the collection of such branch cuts.
The conserved quantities are related to traces of powers of the monodromy
matrix, thus should be insensitive for these branch cuts. As a result,
quasi-momenta may be permuted up to multiples of $2\pi$ on the cuts
\[
q_{\ell}\left(\lambda+i\varepsilon\right)-q_{\ell+1}\left(\lambda-i\varepsilon\right)=2\pi n_{\ell,r},\quad x\in\mathcal{C}_{\ell,r},\:\ell=1,\cdots,\left[N/2\right]-1.
\]
Let us analyze the analytic structure of the quasi-momenta.

\subsubsection{Asymptotics at $\lambda\rightarrow+\infty$}

The large $\lambda$ asymptotics of the transport matrix $T(\lambda)$
can be read off from the definition of the connection (\ref{eq:Jconnection})
\begin{equation}
T\left(\lambda\right)\sim\mathcal{P}\overleftarrow{\exp}\left\{ \frac{1}{\lambda}\int_{0}^{\pi}\mathrm{d\sigma}J_{\tau}\right\} \sim\mathbb{I}+\frac{1}{\lambda}\int_{0}^{\pi}\mathrm{d\sigma}J_{\tau}\equiv\mathbb{I}+\frac{\bar{Q}}{\lambda}+\dots.
\end{equation}
where $\bar{Q}$ would be the conserved charge of the bulk theory.
This implies, for the boundary monodromy matrix: 
\[
\Omega\left(\lambda\right)\sim\mathbb{I}-\frac{1}{\lambda}\left(U\bar{Q}U+\bar{Q}\right)+\dots=\mathbb{I}-\frac{2Q}{\lambda}+\dots\quad;\qquad Q\in\mathfrak{h}\,.
\]
Clearly, only the charges corresponding to the survived symmetry appear
in the asymptotic behaviour of the monodromy matrix. For its eigenvalues,
this implies that 
\[
q_{\ell}\left(\lambda\right)\sim\frac{1}{\lambda}\, q_{\ell}^{[\infty]}+\mathcal{O}\left(\lambda^{-2}\right),
\]
where $q_{l}^{[\infty]}$ are the eigenvalues of the conserved charges
preserved by the boundary.

\subsubsection{Reflection symmetry}

The boundary nature of the monodromy matrix, together with $U^{2}=\mathbb{I},$
implies the following reflection property 
\[
\Omega(\lambda)=U\Omega^{-1}(-\lambda)U\,.
\]
Let us see how this relation translates to the quasi-momenta. Assuming
that $\Omega(\lambda)$ is diagonalized as $\Omega(\lambda)=A(\lambda)\Omega_{\mathrm{diag}}\left(\lambda\right)A(\lambda)^{-1}$
and that the quasi-momenta are all different for generic $\lambda$,
we have 
\[
\Omega_{\mathrm{diag}}\left(\lambda\right)=P\Omega_{\mathrm{diag}}\left(-\lambda\right)^{-1}P^{-1}\quad;\qquad P=A(\lambda)^{-1}UA(-\lambda)
\]
Since $P$ connects a generic diagonal matrix to another generic diagonal
matrix, it has to be a permutation, actually the same permutation
for any $\lambda$. To identify the permutation, we analyze the relation
at $\lambda\to\pm\infty$. In this limit 
\begin{equation}
\Omega_{\mathrm{diag}}\left(\lambda\right)\simeq\mathbb{I}+\frac{2Q}{\lambda}=P\left(\mathbb{I}-\frac{2Q}{-\lambda}\right)P\simeq P\left(\Omega_{\mathrm{diag}}\left(-\lambda\right)\right)^{-1}\, P
\end{equation}
For a generically charged state it implies that $P=1$, which leads
to the following reflection property of the quasi-momenta 
\begin{equation}
q_{\ell}\left(-\lambda\right)=-q_{\ell}\left(\lambda\right).
\end{equation}

\subsubsection{Singularities around $\lambda=\pm1$}

Let us first recall that 
\begin{eqnarray}
\ac_{\sigma}\left(\lambda\right) & = & \frac{1}{1-\lambda^{2}}J_{\sigma}+\frac{\lambda}{1-\lambda^{2}}J_{\tau}=\frac{J_{+}}{1-\lambda}-\frac{J_{-}}{1+\lambda}\label{eq:connection-sigma}
\end{eqnarray}
Thus the pole singularities around $\lambda=\pm1$ are governed by
the light-cone components of the currents $J_{\pm}$. Let us denote
the matrix which diagonalizes these currents by $h_{\pm}\left(\sigma\right)$:
\begin{equation}
J_{\pm}^{\mathrm{diag}}\left(\sigma\right)=h_{\pm}\left(\sigma\right)^{-1}J_{\pm}\left(\sigma\right)h_{\pm}\left(\sigma\right).\label{eq:diag-currents-lambdapm1}
\end{equation}
For the monodromy matrix we have at leading order around $\lambda=\pm1$
\begin{eqnarray}
\left.\Omega\left(\lambda\right)\right|_{\lambda=\pm1} & = & U\, h_{\mp}\left(0\right)\,\exp\left(\frac{1}{\lambda\mp1}\int_{0}^{\pi}\mathrm{d}\sigma J_{\mp}^{\mathrm{diag}}\left(\sigma\right)\right)\, h_{\mp}\left(\pi,\right)^{-1}\, U\times\label{eq:full-monodromy-at-lambdapm1}\\
 &  & \,\,\, h_{\pm}\left(\pi\right)\,\exp\left(\frac{1}{\lambda\mp1}\int_{0}^{\pi}\mathrm{d}\sigma J_{\pm}^{\mathrm{diag}}\left(\sigma\right)\right)\, h_{\pm}\left(0\right)^{-1}.\nonumber 
\end{eqnarray}
Recall from (\ref{eq:Ui-commutation-jpm}) that at the boundaries
the light-cone currents are related by the automorphism, $\alpha,$
as $J_{-}\left(\sigma_{i}\right)=U\, J_{+}\left(\sigma_{i}\right)\, U$,
where $\sigma_{i}=0,\pi$. This means that, at the boundary, the matrices
which diagonalize $J_{-}$ and $J_{+}$ are related, and 
\begin{equation}
J_{\mp}^{\mathrm{diag}}\left(\sigma_{i}\right)=h_{\mp}\left(\sigma_{i}\right)^{-1}Uh_{\pm}\left(\sigma_{i}\right)J_{\pm}^{\mathrm{diag}}\left(\sigma_{i}\right)h_{\pm}\left(\sigma_{i}\right)^{-1}Uh_{\mp}\left(\sigma_{i}\right)\,.
\end{equation}
Actually the currents $J=m\,\dd m=-2(\dd\n)\n^{t}+2\n\,\dd\n^{t}$
are rank $2$ matrices and their diagonal form is 
\[
J_{\mp}^{\mathrm{diag}}\left(\sigma\right)={\rm diag}(i\, j_{\mp}(\sigma),-i\, j_{\mp}(\sigma),0,\dots,0)\,,
\]
thus $h_{\mp}\left(\sigma_{i}\right)^{-1}Uh_{\pm}\left(\sigma_{i}\right)$
is either the identity or the permutation matrix on the relevant 2
dimensional space. Here we assumed that $j_{\mp}(\sigma)$ is never
vanishing, thus the 2 dimensional nonzero subspace is the same for
any $\sigma$. Let us introduce 
\[
\int_{0}^{\pi}\mathrm{d}\sigma j_{\mp}\left(\sigma\right)=\kappa_{\mp},
\]
such that for the quasi-momenta we have \emph{
\begin{eqnarray*}
(q_{1},-q_{1},\dots)\vert_{\lambda\sim1} & = & \frac{\kappa}{\lambda-1}(1,-1,0,\dots,0)+\dots\\
(q_{1},-q_{1},\dots)\vert_{\lambda\sim-1} & = & \frac{\pm\kappa}{\lambda+1}(1,-1,0,\dots,0)+\dots\,,
\end{eqnarray*}
}where $\kappa=\kappa_{+}\pm\kappa_{-}$ depending whether the permutation
is the identity or not. However, consistency with the parity properties
favors the $\kappa=\kappa_{+}+\kappa_{-}$ choice and the plus sign
in the behaviour of $q_{1}$ at $\lambda\sim-1$, i.e. at the boundary
$J_{\mp}^{\mathrm{diag}}\left(\sigma_{i}\right)=J_{\pm}^{\mathrm{diag}}\left(\sigma_{i}\right)$.

\subsubsection{Inversion symmetry}

To derive the behaviour of $\Omega$ under the inversion transformation
we recall that the Lax connection transforms under inversion as 
\[
\ac(\lambda^{-1})=m(d+\ac(\lambda))m\,.
\]
This implies for the transport matrix that 
\[
T(\lambda^{-1})=m(\pi)T(\lambda)m(0),
\]
where $m(0)$ and $m(\pi)$ are the boundary values of the $O(N)$
element. Since in general these values are different for the two different
boundaries, this is not a similarity transformation. Nevertheless,
using it together with $[U,m\vert_{{\rm bdry}}]=0$ in the definition
of $\Omega$ one finds that 
\[
\Omega(\lambda^{-1})=m(0)^{-1}\Omega(\lambda)m(0)\,.
\]
We can now argue similarlyto the case of the reflection symmetry:
the eigenvalues of the monodromy at $\lambda^{-1},$ $\Omega_{\mathrm{diag}}(\lambda^{-1})$,
and of the original monodromy matrix $\Omega_{\mathrm{diag}}(\lambda)$,
are related by a $\lambda$-independent parity transformation:
\begin{equation}
\Omega_{\mathrm{diag}}\left(\lambda\right)=P\Omega_{\mathrm{diag}}\left(\lambda^{-1}\right)P^{-1}\quad;\qquad P=A(\lambda)^{-1}m(0)A(\lambda^{-1})\,.\label{eq:inversionOmegadiag}
\end{equation}
 Analysing the $\lambda\to\lambda^{-1}$ transformation for the singularities
around $\lambda=\pm1$ we can conclude that 
\[
q_{1}(1/\lambda)=-q_{1}(\lambda)\,.
\]
Restricting the relation (\ref{eq:inversionOmegadiag}) for the non-singular
part of the monodromy matrix and evaluating at $\lambda=1$ implies
that the quasi-momenta $q_{j}$ for $j\ne1$ will not suffer any permutations
under inversion:
\begin{equation}
q_{j}\left(1/\lambda\right)=q_{j}\left(\lambda\right).
\end{equation}
Let us note one difference from the bulk theory. When there are no
boundaries, inversion symmetry allows $q_{1}^{\mathrm{bulk}}\left(1/\lambda\right)=2\pi n-q_{1}^{\mathrm{bulk}}\left(\lambda\right)$.
But when we have boundaries, the parity condition enforces this mode
number to be zero.

\subsection{Explicit example in the $O(4)$ model}

In the following, we construct explicitly the spectral curve of a
solution having $SO(2)\times SO(2)$ symmetry. At the language of
the $\n$ variables it corresponds to two Dirichlet and two Neumann
boundary conditions 
\[
\partial_{\sigma}n_{1}\vert=0\quad;\qquad\partial_{\sigma}n_{2}\vert=0\quad;\qquad n_{3}\vert=0\quad;\qquad n_{4}\vert=0\,.
\]
The $U$ matrix in the Lax formulation is 
\[
U={\rm diag}(1,1,-1,-1)\,.
\]
A solution satisfying this boundary condition is 
\[
n_{1}=\cos n\sigma\cos\omega\tau\quad;\qquad n_{2}=\cos n\sigma\sin\omega\tau\quad;\qquad n_{3}=\sin n\sigma\cos\omega\tau\quad;\qquad n_{4}=\sin n\sigma\sin\omega\tau\,.
\]
One can check that both the bulk equations of motion and the boundary
conditions are satisfied, provided $n\in\mathbb{Z}.$ The solution
corresponds to a circular rotating string in $S^{3}$, where the end
points of the string rotate on the same $S^{1}\subset S^{3}$. The
Lax connection, $a_{\sigma}(\lambda)$, is a complicated function
of $\sigma$, making it very difficult to calculate the path-order
exponential. To avoid this problem we switch to the coset formulation.
One possible evolution on the group manifold can be described as 
\[
g(\tau,\sigma)=e^{-n\sigma(J_{13}+J_{24})}e^{-\omega\tau(J_{12}+J_{34})}\quad;\qquad(J_{ik})_{lm}=\delta_{il}\delta_{km}-\delta_{im}\delta_{kl}
\]
Note that we can recover the constrained fields of the circular string
via the relations (\ref{eq:coset-constr}). A gauge equivalent evolution
can be obtained by changing the sign of $J_{34}$. What is nice about
this choice is that $[J_{13}+J_{24},J_{12}+J_{34}]=0$, and the components
of the Maurer-Cartan one-form are constants: 
\[
\omega_{\sigma}=g^{-1}\partial_{\sigma}g=-n(J_{13}+J_{24})\quad;\qquad\omega_{\tau}=g^{-1}\partial_{\tau}g=-\omega(J_{12}+J_{34})\,.
\]
The projected currents are 
\[
A_{\sigma}=-nJ_{24}\quad;\qquad A_{\tau}=-\omega J_{34}\quad;\qquad K_{\sigma}=-nJ_{13}\quad;\qquad K_{\tau}=-\omega J_{12}\,,
\]
which satisfy the equation of motion (\ref{eq:EOM}). The $\sigma$
component of the Lax connection is
\[
\Lc_{\sigma}(\lambda)=-nJ_{24}-n\frac{\lambda^{2}+1}{\lambda^{2}-1}J_{13}-\omega\frac{2\lambda}{\lambda^{2}-1}J_{12}\,.
\]
Since 
\[
-U\Lc_{\sigma}(-\lambda)U=\Lc_{\sigma}(\lambda)\quad\longrightarrow\quad UT(-\lambda)^{-1}U=T(\lambda)
\]
we can simply diagonalize $\Lc_{\sigma}(\lambda)$ and exponentiate
it to get the eigenvalues of the monodromy matrix. We find 
\[
q_{1,2}=\frac{2\pi}{1-\lambda^{2}}i\sqrt{n^{2}(\lambda^{4}+1)+2\omega^{2}\lambda^{2}\pm2A}\quad;\qquad A=\lambda^{2}\sqrt{(\omega^{2}+n^{2}\lambda^{2})(n^{2}/\lambda^{2}+\omega^{2})}\,.
\]
It is instructive to write the quasi-momenta as 
\[
q_{1}=\frac{2\pi i\lambda}{1-\lambda^{2}}\left(\sqrt{\omega^{2}+n^{2}\lambda^{2}}+\sqrt{\frac{n^{2}}{\lambda^{2}}+\omega^{2}}\right)\quad;\qquad q_{2}=\frac{2\pi i\lambda}{1-\lambda^{2}}\left(\sqrt{\omega^{2}+n^{2}\lambda^{2}}-\sqrt{\frac{n^{2}}{\lambda^{2}}+\omega^{2}}\right).
\]
This solution is analogous to the open string solution of the $Y=0$
brane \cite{Bajnok:2013sza}. The point-like string solution does
not depend on $\sigma$ and corresponds to $n=0$. Its quasi-momenta
are 
\[
q_{1}=\frac{4\pi i\lambda\omega}{1-\lambda^{2}}\quad;\qquad q_{2}=0
\]
One can easily check that the spectral curve has the right asymptotics
and residues around $\lambda=\pm1$, and satisfies the inversion and
reflection properties. Let us also note that the point-like string
solution satisfies the all Neumann boundary condition, too.

\section{Quantum integrability\label{sec:Quantum-integrability}}

The quantum integrability of the $O(N)$ non-linear $\sigma$-models
can be shown by following the argumentations of Polyakov \cite{Polyakov:1977vm},
Goldschmidt and Witten (GW) \cite{Goldschmidt:1980wq}. The idea is
to analyze the classical conservation laws (\ref{eq:cl_cons_law})
and their possible quantum corrections. Since products of operators
are not well-defined at the quantum level they have to be regularized,
leading to the appearance of new terms which can make the classical
symmetry anomalous. To decide whether a higher spin symmetry is maintained
or not, one has to classify the possible anomaly terms. In the case
of $T_{++}^{2}$, the global symmetry, the dimensionality of the fields
and the Lorentz transformation property fix the anomaly of the form
\[
\partial_{-}T_{++}^{2}=c_{1}\partial_{-}\left(\partial_{+}^{2}\n^{t}\partial_{+}^{2}\n\right)+c_{2}\partial_{+}\left(\partial_{-}\n^{t}\partial_{+}\n\partial_{+}\n^{t}\partial_{+}\n\right)+c_{3}\partial_{+}\left(\partial_{+}^{3}\n^{t}\partial_{-}\n\right)\,.
\]
Grouping the terms a conservation law can be established at the quantum
level -- of the form (\ref{eq:Bulk_cons}) --

\begin{equation}
\partial_{-}T^{(4)}=\partial_{+}\Theta^{(2)}\label{eq:Q_cons}
\end{equation}
with 
\[
T=T_{++}^{2}-c_{1}\left(\partial_{+}^{2}\n^{t}\partial_{+}^{2}\n\right)\quad;\qquad\Theta^{(2)}=c_{2}\left(\partial_{-}\n^{t}\partial_{+}\n\partial_{+}\n^{t}\partial_{+}\n\right)+c_{3}\left(\partial_{+}^{3}\n^{t}\partial_{-}\n\right),
\]
 which implies factorized scattering. We can obtain a similar equation
by exchanging $\partial_{+}\leftrightarrow\partial_{-}$. 

Integrability in the presence of boundaries, similarly to the classical
case, requires the fulfillment of the equation 
\[
\left[T^{(s+1)}-\bar{T}^{(s+1)}+\bar{\Theta}^{(s-1)}-\Theta^{(s-1)}\right]\vert=\frac{d\Sigma}{dt}\,.
\]
Since at the quantum level the bulk-boundary OPEs can get quantum
corrections, classical integrable boundary conditions can be anomalous.
In particular, for the $O(N)$ symmetric Neumann boundary condition,
$\partial_{1}\n=0$, symmetry and dimensionality allow for an anomaly
of the form 
\[
T_{++}-T_{--}\vert=c\partial_{\tau}\n^{t}\partial_{\tau}\n\,,
\]
which would even spoil the existence of a conserved energy. Since
we have no control of such terms, nor have we a systematic quantization
approach which can decide in these questions, we classify the integrable
boundary conditions based on the existence of reflection factors,
which satisfy the boundary bootstrap equations.

\subsection{Reflection factors}

As we explained, the GW argument implies that the $O(N)$ non-linear
$\sigma$-model is integrable at the quantum level. There are $N$
particles with the same mass, $m$, and they transform in the vector
representation of $O(N)$. The index structure of the scattering matrix,
$\mathcal{S}$, compatible with this symmetry, has the form 
\[
S_{ij}^{kj}(\theta)=\sigma_{1}(\theta)\delta_{ij}\delta^{kl}+\sigma_{2}(\theta)\delta_{i}^{k}\delta_{j}^{l}+\sigma_{3}(\theta)\delta_{i}^{l}\delta_{j}^{k}\,.
\]
Factorized scattering implies the Yang-Baxter equation (YBE) which,
together with unitarity and crossing symmetry, restricts the amplitudes
to
\[
\sigma_{1}(\theta)=-\frac{i\lambda}{i\pi-\theta}\sigma_{2}(\theta)\quad;\qquad\sigma_{3}(\theta)=-\frac{i\lambda}{\theta}\sigma_{2}(\theta)\quad;\qquad\lambda=\frac{2\pi}{N-2}\,,
\]
where 
\[
\sigma_{2}(\theta)=\frac{\Gamma(\frac{1}{2}+\frac{\lambda}{2\pi}+\frac{i\theta}{2\pi})}{\Gamma(\frac{1}{2}+\frac{\lambda}{2\pi}-\frac{i\theta}{2\pi})}\frac{\Gamma(1+\frac{i\theta}{2\pi})}{\Gamma(-\frac{i\theta}{2\pi})}\frac{\Gamma(\frac{1}{2}-\frac{i\theta}{2\pi})}{\Gamma(\frac{1}{2}+\frac{i\theta}{2\pi})}\frac{\Gamma(\frac{\lambda}{2\pi}-\frac{i\theta}{2\pi})}{\Gamma(1+\frac{\lambda}{2\pi}+\frac{i\theta}{2\pi})}
\]

In the presence of an integrable boundary particles reflect from the
boundary by a reflection matrix $R_{i}^{j}(\theta)$, which satisfies
the boundary analogue of the Yang-Baxter equation
\[
S_{ji}^{nm}(\theta_{1}-\theta_{2})R_{n}^{p}(\theta_{1})S_{mp}^{ql}(\theta_{1}+\theta_{2})R_{q}^{k}(\theta_{2})=R_{i}^{p}(\theta_{2})S_{jp}^{nm}(\theta_{1}+\theta_{2})R_{n}^{q}(\theta_{1})S_{mq}^{kl}(\theta_{1}-\theta_{2}).
\]
The scalar factor is fixed from unitarity and boundary crossing unitarity
\cite{Ghoshal:1993tm}
\[
R_{i}^{k}(\theta)R_{k}^{j}(-\theta)=\delta_{i}^{j}\qquad;\qquad R_{i}^{j}(\frac{i\pi}{2}-\theta)=S_{kl}^{ij}(2\theta)R_{k}^{l}(\frac{i\pi}{2}+\theta).
\]
The solutions of these equations were classified and they fall into
the following classes: 
\begin{enumerate}
\item \textbf{Diagonal} \cite{Moriconi:1998gc}. The diagonal reflection
factors have the form 
\[
\mathcal{R}(\theta)={\rm diag}(\underbrace{R_{1}(\theta),\dots,R_{1}(\theta)}_{k},\underbrace{R_{2}(\theta),\dots,R_{2}(\theta)}_{N-k})
\]
where from the boundary Yang-Baxter equation it follows that
\[
\frac{R_{2}(\theta)}{R_{1}(\theta)}=\frac{c-\theta}{c+\theta}\quad;\qquad c=-i\frac{\pi}{2}\frac{N-2k}{N-2}.
\]
Clearly, for the transformation $k\leftrightarrow N-k$ the reflection
factors exchange $R_{1}\leftrightarrow R_{2}$, thus it is enough
to consider the cases $k\leq N/2$. Unitarity and boundary crossing
unitarity fix the scalar factor up to a CDD factor as 
\begin{eqnarray*}
R_{1}(\theta) & = & -R_{0}(\theta)\frac{\Gamma(\frac{1}{4}+\frac{\lambda}{4\pi}+\frac{i\theta}{2\pi})}{\Gamma(\frac{1}{4}+\frac{\lambda}{4\pi}-\frac{i\theta}{2\pi})}\frac{\Gamma(\frac{3}{4}+\frac{\lambda}{4\pi}-\frac{i\theta}{2\pi})}{\Gamma(\frac{3}{4}+\frac{\lambda}{4\pi}+\frac{i\theta}{2\pi})}\frac{\Gamma(\frac{1}{4}+\frac{\lambda(N-k-1)}{4\pi}+\frac{i\theta}{2\pi})}{\Gamma(\frac{1}{4}+\frac{\lambda(N-k-1)}{4\pi}-\frac{i\theta}{2\pi})}\frac{\Gamma(\frac{3}{4}+\frac{\lambda(N-k-1)}{4\pi}-\frac{i\theta}{2\pi})}{\Gamma(\frac{3}{4}+\frac{\lambda(N-k-1)}{4\pi}+\frac{i\theta}{2\pi})};\\
R_{0}(\theta) & = & \frac{\Gamma(\frac{1}{2}+\frac{\lambda}{4\pi}-\frac{i\theta}{2\pi})}{\Gamma(\frac{1}{2}+\frac{\lambda}{4\pi}+\frac{i\theta}{2\pi})}\frac{\Gamma(1+\frac{i\theta}{2\pi})}{\Gamma(1-\frac{i\theta}{2\pi})}\frac{\Gamma(\frac{1}{4}-\frac{i\theta}{2\pi})}{\Gamma(\frac{1}{4}+\frac{i\theta}{2\pi})}\frac{\Gamma(\frac{3}{4}+\frac{\lambda}{4\pi}+\frac{i\theta}{2\pi})}{\Gamma(\frac{3}{4}+\frac{\lambda}{4\pi}-\frac{i\theta}{2\pi})}.
\end{eqnarray*}
The symmetry of this boundary condition is $SO(k)\times SO(N-k)$.
The case $k=N-1$ corresponds to the fixed boundary condition of Ghoshal
\cite{Ghoshal:1994bc}, i.e. all boundary conditions are Dirichlet,
while the case $k=0$ corresponds to the $O(N)$ symmetric Neumann
boundary condition of Ghoshal \cite{Ghoshal:1994bc}. Finally, we
also note that in the $\theta\to\infty$ limit the reflection factor
agrees with the boundary Lax matrix $U$. 
\item \textbf{One-block} \cite{Moriconi:2001xz}. The reflection factor
is diagonal except a $2\times2$ block
\[
\mathcal{R}(\theta)=\left(\begin{array}{ccccc}
A_{\xi}(\theta) & B_{\xi}(\theta) & 0 & \cdots & 0\\
-B_{\xi}(\theta) & A_{\xi}(\theta) & 0 & \cdots & 0\\
0 & 0 & R_{\xi}(\theta) & \cdots & 0\\
\vdots & \vdots & \vdots & \ddots & \vdots\\
0 & 0 & 0 & \cdots & R_{\xi}(\theta)
\end{array}\right)
\]
where 
\[
A(\theta)=\frac{1}{2}\left(\frac{c-\theta}{c+\theta}+\frac{c'-\theta}{c'+\theta}\right)R(\theta)\quad;\qquad B(\theta)=\frac{1}{2i}\left(\frac{c-\theta}{c+\theta}-\frac{c'-\theta}{c'+\theta}\right)R(\theta)\,,
\]
with the constraint 
\[
c=-\frac{i\pi}{2}\frac{N-4}{N-2}+\xi\quad;\qquad c'=-\frac{i\pi}{2}\frac{N-4}{N-2}-\xi.
\]
The symmetry of this boundary condition is $SO(2)\times SO(N-2)$.
The case $\xi=0$ reduces to the diagonal solution with $k=2$ above.
There is another diagonal limit of the reflection factor, namely by
sending $\xi\to\infty$ we can recover the $O(N)$ symmetric boundary
condition. Unitarity, together with boundary crossing unitarity, fixes
the reflection factors to be:
\begin{eqnarray*}
R_{\xi}(\theta) & = & -R_{0}(\theta)\frac{\Gamma(\frac{1}{4}+\frac{\lambda+2i\xi}{4\pi}+\frac{i\theta}{2\pi})}{\Gamma(\frac{1}{4}+\frac{\lambda+2i\xi}{4\pi}-\frac{i\theta}{2\pi})}\frac{\Gamma(\frac{3}{4}+\frac{\lambda+2i\xi}{4\pi}-\frac{i\theta}{2\pi})}{\Gamma(\frac{3}{4}+\frac{\lambda+2i\xi}{4\pi}+\frac{i\theta}{2\pi})}\frac{\Gamma(\frac{1}{4}+\frac{\lambda-2i\xi}{4\pi}+\frac{i\theta}{2\pi})}{\Gamma(\frac{1}{4}+\frac{\lambda-2i\xi}{4\pi}-\frac{i\theta}{2\pi})}\frac{\Gamma(\frac{3}{4}+\frac{\lambda-2i\xi}{4\pi}-\frac{i\theta}{2\pi})}{\Gamma(\frac{3}{4}+\frac{\lambda-2i\xi}{4\pi}+\frac{i\theta}{2\pi})}\,.
\end{eqnarray*}

\item \textbf{All blocks the same} \cite{Moriconi:2001xz}. There is also
a completely non-diagonal reflection factor for any even $N$ of the
form 
\[
\mathcal{R}(\theta)=\left(\begin{array}{ccccc}
A(\theta) & B(\theta) & 0 & 0 & \cdots\\
-B(\theta) & A(\theta) & 0 & 0 & \cdots\\
0 & 0 & A(\theta) & B(\theta) & \cdots\\
0 & 0 & -B(\theta) & A(\theta) & \cdots\\
\vdots & \vdots & \vdots & \vdots & \ddots
\end{array}\right).
\]
From the boundary YBE it follows that 
\[
B(\theta)=\alpha\,\theta\, A(\theta)\,.
\]
The symmetry of the boundary condition is $U(N/2)$. Unitarity and
boundary crossing unitarity fix the scalar factor to 
\[
A(\theta)=-\frac{1}{2\pi\alpha}\frac{\Gamma\left(\frac{1}{2}-\frac{1}{2\pi\alpha}+\frac{i\theta}{2\pi}\right)\Gamma\left(-\frac{1}{2\pi\alpha}-\frac{i\theta}{2\pi}\right)}{\Gamma\left(\frac{1}{2}-\frac{1}{2\pi\alpha}-\frac{i\theta}{2\pi}\right)\Gamma\left(1-\frac{1}{2\pi\alpha}+\frac{i\theta}{2\pi}\right)}R_{0}(\theta)\,.
\]

\item \textbf{Exceptional} boundary conditions for the $O(4)$ model. The
above classification was confirmed in \cite{Arnaudon:2003gj}, and
additionally a new family of boundary conditions was found in the
$O(4)$ model. Since $\SO{4}\equiv\SU{2}_{l}\times\SU{2}_{r}$, the
$O(4)$ model is an $SU(2)$ principal chiral model, which allows
a two parameter, $(\xi_{l},\xi_{r})$~family of reflection factors:
\begin{align*}
\mathcal{R}(\theta)=\begin{pmatrix}A_{+}(\theta) & B_{+}(\theta) & 0 & 0\\
-B_{+}(\theta) & A_{+}(\theta) & 0 & 0\\
0 & 0 & A_{-}(\theta) & B_{-}(\theta)\\
0 & 0 & -B_{-}(\theta) & A_{-}(\theta)
\end{pmatrix},
\end{align*}
where 
\[
A_{\pm}(\theta)=R(\theta)\frac{\xi_{l}\xi_{r}\pm\theta^{2}}{(\xi_{l}+\theta)(\xi_{r}+\theta)}\qquad;\qquad B_{\pm}(\theta)=\frac{-i\theta(\xi_{r}\pm\xi_{l})}{(\xi_{l}+\theta)(\xi_{r}+\theta)}
\]
and 
\[
R(\theta)=\frac{\Gamma\left(\frac{1}{2}-\frac{i\xi_{l}}{2\pi}+\frac{i\theta}{2\pi}\right)}{\Gamma\left(\frac{1}{2}-\frac{i\xi_{l}}{2\pi}-\frac{i\theta}{2\pi}\right)}\frac{\Gamma\left(1-\frac{i\xi_{l}}{2\pi}-\frac{i\theta}{2\pi}\right)}{\Gamma\left(1-\frac{i\xi_{l}}{2\pi}+\frac{i\theta}{2\pi}\right)}\frac{\Gamma\left(\frac{1}{2}-\frac{i\xi_{r}}{2\pi}+\frac{i\theta}{2\pi}\right)}{\Gamma\left(\frac{1}{2}-\frac{i\xi_{r}}{2\pi}-\frac{i\theta}{2\pi}\right)}\frac{\Gamma\left(1-\frac{i\xi_{r}}{2\pi}-\frac{i\theta}{2\pi}\right)}{\Gamma\left(1-\frac{i\xi_{r}}{2\pi}+\frac{i\theta}{2\pi}\right)}R_{0}(\theta)\,.
\]
The symmetry of this boundary condition is $U(1)_{l}\times U(1)_{r}$.
By choosing $\xi_{l}=-\xi_{r}$, the solution reduces to the one-block
case, while choosing $\xi_{r}=i/\alpha$ and taking the $\xi_{l}\rightarrow\infty$
limit we can recover the two block, same reflection factor.
\end{enumerate}
Comparing this classification with the classical case, we can see
that the ``all block different'' boundary reflection factor exists
only in the $O(4)$ case. It would be interesting to understand how
the general integrable boundary conditions, formulated in terms of
the antisymmetric matrix $M$, become anomalous during the quantization
procedure. From the existence of the nondiagonal reflection factors
in the quantum case it is expected that they also have classical limits,
which should be described by monodromy matrices and spectral curves.
In order to make a connection to the classical formulations, we analyze
the classical limit of the spectrum via the Bethe-Yang equations.

\subsection{Bethe-Yang equations }

In this section we analyze the large volume spectrum in a finite volume.
The energy of an $n$-particle state on an interval of size $L$,
with rapidities $\theta_{1},\dots,\theta_{n}$, can be written as
\begin{equation}
E=\sum_{i=1}^{n}m\cosh\theta_{i}+O\left(e^{-mL}\right)\,.
\end{equation}
Volume dependence comes through momentum quantisation, which formulates
the periodicity of the wave function. The requirement is called the
Bethe-Yang equation, which is based on the infinite volume scattering
and reflection matrices as
\begin{equation}
e^{im2L\sinh\theta_{i}}\prod_{j=i+1}^{n}\mathcal{S}(\theta_{i}-\theta_{j})\mathcal{R}(\theta_{i})\prod_{j=n:j\neq i}^{1}\mathcal{S}(\theta_{j}+\theta_{i})\mathcal{R}(-\theta_{i})\prod_{j=1}^{i-1}\mathcal{S}(\theta_{i}-\theta_{j})=1,\label{eq:BBY}
\end{equation}
where we assumed that the left and the right boundaries are the same.
These matrix equations can be solved by diagonalizing the double row
transfer matrix 
\begin{equation}
T(\theta_{0}\vert\{\theta_{i}\})=\mbox{tr}_{0}\left(\prod_{j=1}^{n}\mathcal{S}(\theta_{0}-\theta_{j})\mathcal{R}(\theta_{0})\prod_{j=n}^{1}\mathcal{S}(\theta_{j}+\theta_{0})\mathcal{R}^{c}(-\theta_{0})\right),
\end{equation}
where the charge conjugation of the reflection factor ensures the
equivalence to eq. (\ref{eq:BBY}) (see \cite{Bajnok:2010ui} for
details). The diagonalization can be done either via the analytic
\cite{Arnaudon:2003gj} or the algebraic \cite{Gombor:2015kdu} Bethe
Ansatz (BA). Results are available only for even $N$ so we restrict
ourselves to those cases. In the analytic BA, the regularity of the
transfer matrix at the positions of the Bethe roots provides the BA
equations. They can be most compactly described using roots and functions
related to the extended Dynkin diagram for $O(N)$. The original Dynkin
diagram of $D_{N/2}$ is extended by a black dot, with label $0$,
which represents the massive particle, see Figure \ref{Dynkin}. Additionally
there are magnonic particles with labels, $i=1,\dots,N/2-2,+,-$.
For each index we associate two $Q$-functions:

\begin{figure}
\begin{centering}
\includegraphics[width=6cm]{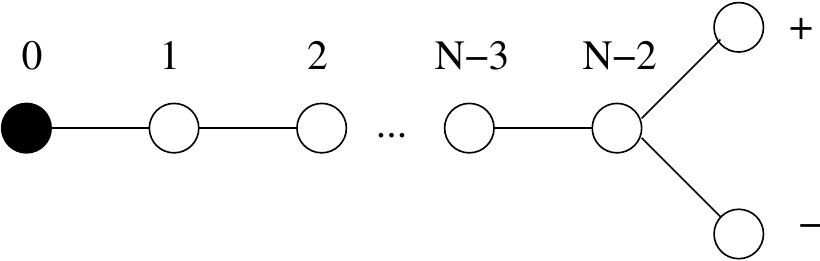}\label{Dynkin}
\par\end{centering}

\protect\caption{Extended Dynkin diagram of $O(N)$ for even $N$. }
\end{figure}

\begin{align}
Q_{i}(\theta) & =\prod_{k=1}^{n_{i}}(\theta-\ui{i}{k})(\theta+\ui{i}{k}),\\
\tilde{Q}_{i}(\theta) & =\prod_{\substack{k=1\\
\theta\neq\ui{i}{k}
}
}^{n_{i}}(\theta-\ui{i}{k})(\theta+\ui{i}{k}).
\end{align}
The one with tilde leaves out the root, if it is evaluated at the
root position. All the magnonic BA equations can be formulated very
compactly as: 
\begin{align}
\frac{Q_{i-1}^{-}(\ui{i}{k})}{Q_{i-1}^{+}(\ui{i}{k})}\frac{\tilde{Q}_{i}^{++}(\ui{i}{k})}{\tilde{Q}_{i}^{--}(\ui{i}{k})}\frac{Q_{i+1}^{-}(\ui{i}{k})}{Q_{i+1}^{+}(\ui{i}{k})} & =r_{i}^{0}(\ui{i}{k})r_{i}^{\pi}(\ui{i}{k}),\qquad0<i<\frac{N}{2}-2,\\
\frac{Q_{-}^{-}(\ui{i}{k})}{Q_{-}^{+}(\ui{i}{k})}\frac{\tilde{Q}_{i}^{++}(\ui{i}{k})}{\tilde{Q}_{i}^{--}(\ui{i}{k})}\frac{Q_{+}^{-}(\ui{i}{k})}{Q_{+}^{+}(\ui{i}{k})} & =r_{i}^{0}(\ui{i}{k})r_{i}^{\pi}(\ui{i}{k}),\qquad i=\frac{N}{2}-2,\\
\frac{Q_{N/2-2}^{-}(\ui{-}{k})}{Q_{N/2-2}^{+}(\ui{-}{k})}\frac{\tilde{Q}_{-}^{++}(\ui{-}{k})}{\tilde{Q}_{-}^{--}(\ui{-}{k})} & =r_{-}^{0}(\ui{-}{k})r_{-}^{\pi}(\ui{-}{k}),\\
\frac{Q_{N/2-2}^{-}(\ui{+}{k})}{Q_{N/2-2}^{-}(\ui{+}{k})}\frac{\tilde{Q}_{+}^{++}(\ui{+}{k})}{\tilde{Q}_{+}^{--}(\ui{+}{k})} & =r_{+}^{0}(\ui{+}{k})r_{+}^{\pi}(\ui{+}{k}),
\end{align}
where $f^{\pm}(\theta)=f(\theta\pm\frac{i\pi}{N-2})$ and the BA equations
for the massive particles reads as: 
\begin{align*}
e^{i2p_{k}L}\prod_{\substack{j=1\\
j\neq k
}
}^{n_{0}}S(\ui{0}{k}-\ui{0}{j})\frac{Q_{1}^{-}(\ui{0}{k})}{Q_{1}^{+}(\ui{0}{k})} & =r_{0}^{0}(\ui{i}{k})r_{0}^{\pi}(\ui{i}{k}).
\end{align*}
Above, we introduced the function 
\[
S(\theta)=\frac{\Gamma(-\frac{i\theta}{2\pi})}{\Gamma(1-\frac{i\theta}{2\pi})}\sigma_{2}(\theta).
\]
The dependence on the boundary conditions sits in the various reflection
phases, of which the non-trivial ones ($\neq1$) are the following: 
\begin{enumerate}
\item $\SO{k}\times\SO{N-k}$ case with even $k$ ($k\le N/2$): 
\begin{align}
r_{0}(\theta) & =R_{1}(\theta),\\
r_{i}(\theta) & =\frac{i\frac{\pi}{2}\frac{N-k}{N-2}-\theta}{i\frac{\pi}{2}\frac{N-k}{N-2}+\theta}\delta_{i,k/2}.
\end{align}

\item $\SO{2}\times\SO{N-2}$ case: 
\begin{align}
r_{0}(\theta) & =A(\theta)+iB(\theta),\\
r_{i}(\theta) & =\frac{i\frac{\pi}{2}-\xi-\theta}{i\frac{\pi}{2}-\xi+\theta}\delta_{i,1}.
\end{align}

\item $\Un{N/2}$ case: 
\begin{align}
r_{0}(\theta) & =A(\theta)+iB(\theta),\\
r_{i}(\theta) & =\frac{i\frac{\pi}{2}+\frac{i}{\alpha}-\theta}{i\frac{\pi}{2}+\frac{i}{\alpha}+\theta}\delta_{i,+}.
\end{align}

\item $\Un{1}_{l}\times\Un{1}_{r}$ case: 
\begin{align}
r_{0}(\theta) & =R(\theta),\\
r_{-}(\theta) & =\frac{i\frac{\pi}{2}+\xi_{l}-\theta}{i\frac{\pi}{2}+\xi_{l}+\theta}\delta_{i,-},\\
r_{+}(\theta) & =\frac{i\frac{\pi}{2}+\xi_{r}-\theta}{i\frac{\pi}{2}+\xi_{r}+\theta}\delta_{i,+}.
\end{align}

\end{enumerate}

\subsection{Spectral curve as the limit of the BY equations}

In this section we investigate the (quasi-) classical limit of the
boundary Bethe-Yang equations in the $O(4)$ model, for two boundary
conditions: the free boundary conditions (all four Neumann) and the
the mixed diagonal reflection with two Neumann and two Dirichlet ones. 

The quantum $O(4)$ $\sigma$-model is asymptotically free with a
dynamically generated mass $m=\Lambda\, e^{-\frac{\sqrt{\lambda}}{2}}$
($\Lambda$ being the cutoff and $\lambda$ is the 't Hooft coupling
evaluated at $\Lambda$). Similarly to the periodic case \cite{Gromov:2006dh},
we can only compare this model to a quantum field theory defined by
a Lagrangian in the classical limit $\lambda\rightarrow\infty$ ($m\rightarrow0$),
in which case it exhibits the classical conformal symmetry of the
latter. In this limit, the dimensionless parameter $\mu=mL=\Lambda L\, e^{-\frac{\sqrt{\lambda}}{2}}$
appearing in the BA equations tends to zero, $\mu\rightarrow0$. The
quasi-momenta of the corresponding classical spectral curve then have
cuts, originated from the condensation of Bethe roots, see \cite{Kazakov:2004qf}
for details. Indeed, if we take both the number of particles $n_{0}$
and the number of roots $2n_{+}\sim2n_{-}\sim2n_{0}$ to infinity
($n_{0}\rightarrow\infty$), while keeping the quantization number
fixed, it implies that all roots $\pm\theta_{\alpha}$, $\pm u_{j}$
and $\pm v_{j}$-s become large together with their differences. In
an appropriately rescaled variable, relevant for the classical limit,
they start to condense on cuts. If one introduces the densities of
these condensed roots, then the Bethe ansatz equations provide integral
equations restricting these densities. Finally, the solutions for
the resolvent of the densities can be mapped to the quasi-momenta
of the classical spectral curve. Let us see, how this can be achieved
in the simplest case. 

The Bethe-Yang equations for the roots $\theta_{\beta}=u_{\beta}^{(0)}$
take the following form: 
\begin{equation}
e^{2ip_{\beta}L}\prod\limits _{\alpha:\alpha\neq\beta}^{n_{0}}S_{0}^{2}(\theta_{\beta}+\theta_{\alpha})S_{0}^{2}(\theta_{\beta}-\theta_{\alpha})r_{i}^{2}(\theta_{\beta})\frac{Q_{+}^{+}(\theta_{\beta})}{Q_{+}^{-}(\theta_{\beta})}\frac{Q_{-}^{+}(\theta_{\beta})}{Q_{-}^{-}(\theta_{\beta})}=1,\qquad i=f,m\quad\beta=1,\dots,n_{0}\label{midnode}
\end{equation}
where
\[
S_{0}(\theta)=i\frac{\Gamma(\frac{1}{2}-\frac{i\theta}{2\pi})}{\Gamma(\frac{1}{2}+\frac{i\theta}{2\pi})}\frac{\Gamma(\frac{i\theta}{2\pi})}{\Gamma(-\frac{i\theta}{2\pi})}
\]
and 
\[
r_{f}(\theta)=\frac{\Gamma(\frac{3}{4}-\frac{i\theta}{2\pi})}{\Gamma(\frac{3}{4}+\frac{i\theta}{2\pi})}\frac{\Gamma(1+\frac{i\theta}{2\pi})^{2}}{\Gamma(1-\frac{i\theta}{2\pi})^{2}}\frac{\Gamma(\frac{1}{4}-\frac{i\theta}{2\pi})}{\Gamma(\frac{1}{4}+\frac{i\theta}{2\pi})}\quad;\qquad r_{m}(\theta)=\frac{\Gamma(\frac{3}{4}-\frac{i\theta}{2\pi})}{\Gamma(\frac{3}{4}+\frac{i\theta}{2\pi})}\frac{\Gamma(\frac{1}{2}+\frac{i\theta}{2\pi})^{2}}{\Gamma(\frac{1}{2}-\frac{i\theta}{2\pi})^{2}}\frac{\Gamma(\frac{1}{4}-\frac{i\theta}{2\pi})}{\Gamma(\frac{1}{4}+\frac{i\theta}{2\pi})}\,.
\]
The accompanying Bethe equations for the roots $u_{j}^{(+)}\equiv u_{j}$
are 
\begin{equation}
1=\frac{Q_{0}^{-}(u_{j})}{Q_{0}^{+}(u_{j})}\frac{\tilde{Q}_{+}^{++}(u_{j})}{\tilde{Q}_{+}^{--}(u_{j})},\qquad j=1,\dots,n^{+}\label{bethe1}
\end{equation}
for the free case, while for the mixed case they change to 
\begin{equation}
\Bigl(\frac{u_{j}-\frac{i\pi}{2}}{u_{j}+\frac{i\pi}{2}}\Bigr)^{2}=\frac{Q_{0}^{-}(u_{j})}{Q_{0}^{+}(u_{j})}\frac{\tilde{Q}_{+}^{++}(u_{j})}{\tilde{Q}_{+}^{--}(u_{j})},\qquad j=1,\dots,n^{+}\,.\label{bethe2}
\end{equation}
(The equations for the $u_{j}^{(-)}$ roots, denoted here by $v_{j}$,
are obtained from eqs. (\ref{bethe1}), (\ref{bethe2}) by the $(u_{j},\ n^{+})\rightarrow(v_{j},\ n^{-})$
substitutions). It is straightforward to show, using the transformation
properties of the various functions appearing in eq. (\ref{midnode}),
that if $\theta_{\beta}$ is a solution, then so is $-\theta_{\beta}$.
Note that substituting $u_{j}\rightarrow-u_{j}$ (but keeping $u_{i}$
for $i\neq j$ the same) changes eqs. (\ref{bethe1}), (\ref{bethe2})
to their inverses, thus the roots are also doubled: to every root
$u_{j}$ solving eqs. (\ref{midnode})-(\ref{bethe1}) (or eqs. (\ref{midnode})-(\ref{bethe2}))
there is another one $-u_{j}$.

Apart from the $r_{i}^{2}(\theta_{\beta})$ factors in eq. (\ref{midnode})
and the l.h.s. in eq. (\ref{bethe2}) the system consisting of (\ref{midnode})-(
\ref{bethe1}) (or (\ref{midnode})-(\ref{bethe2})) is identical
to the system of BA equations in the periodic case \cite{Gromov:2006dh}
but for $2n_{0}$ particles with rapidities coming in pairs $(\theta_{\beta},-\theta_{\beta})$,
accompanied by $2n_{+}$ ``left'' roots $(u_{j},-u_{j})$ ($2n_{+}$
``right'' roots $(v_{j},-v_{j})$ ) coming also in pairs. Thus the
effect of the integrability preserving boundaries is twofold: on the
one hand they double the particles and the left/right roots, while
on the other hand they introduce the $r_{i}^{2}(\theta_{\beta})$
factors and the l.h.s. in eq. (\ref{bethe2}). However, in the light
of the observations in the previous paragraph, even in the presence
of these modifications the solutions of the boundary BA equations
above are $n_{0}$ pairs of $(\theta_{\beta},-\theta_{\beta})$, accompanied
by $2n_{+}$ ($2n_{-}$) pairs of roots $(u_{j},-u_{j})$ ($(v_{j},-v_{j})$).

Let us now consider the limit classical limit when $\mu\rightarrow0$.
We also let both the number of particles $n_{0}$, and the number
of roots $2n_{+}\sim2n_{-}\sim2n_{0}$ go to infinity enforcing that
all $\pm\theta_{\alpha}$, $\pm u_{j}$ and $\pm v_{j}$-s become
large allowing us to take the logarithms of the Bethe equations and
use the ``Coulomb approximations'' (large $\theta$ or $u$): 
\[
-i\log S^{2}(\theta)\rightarrow-\frac{\pi}{\theta},\qquad-i\log r_{1}^{2}(\theta)\rightarrow-\frac{3\pi}{2\theta},\qquad-i\log r_{2}^{2}(\theta)\rightarrow\frac{\pi}{2\theta},\qquad\log\frac{u+iB}{u-iB}\rightarrow\frac{2iB}{u}\,.
\]
We consider here the classical limit of eq.(\ref{midnode}) in the
absence of any roots. Then, after taking the log of both sides and
using the previous approximations, we find: 
\[
\frac{2\mu}{\pi}\sinh(\theta_{\beta})-\sum\limits _{\alpha\neq\beta}\Bigl(\frac{1}{\theta_{\beta}+\theta_{\alpha}}+\frac{1}{\theta_{\beta}-\theta_{\alpha}}\Bigr)+{-\frac{3}{2\theta_{\beta}}\atop \ \frac{1}{2\theta_{\beta}}}=2m_{\beta}\,,\quad m_{\beta}\in\mathrm{Z}\quad\beta=1,\dots,n_{0}.
\]
where the upper line applies for the free and the lower line for the
mixed diagonal boundary conditions.

These equations describe a system of (1D) particles put into the combination
of constant external forces ($2m_{\beta}$) and a (confining) potential
$V(\theta)=\frac{2\mu}{\pi}\cosh(\theta)$, which interact by the
Coulomb repulsion not only with each other but also with their ``mirror
images'' at $-\theta_{\beta}$. The interaction of the particles
with the boundaries is described by the last terms. To emphasize our
interpretation we rewrite 
\[
{-\frac{3}{2\theta_{\beta}}\atop \ \frac{}{2\theta_{\beta}}}=-\frac{1}{2\theta_{\beta}}+{-\frac{1}{\theta_{\beta}}\atop \ \frac{1}{\theta_{\beta}}}
\]
since the $1/(2\theta_{\beta})$ term reproduces the Coulomb repulsion
between the $\beta$-th particle and its mirror image (and thus can
be absorbed into the first term of the sum), and the interactions
with the boundaries differs only by a sign between the free and the
mixed diagonal cases. This interaction with the mirror images and
with the boundaries are the two new phenomena which are absent in
the periodic case.

It is convenient to handle the asymptotically large nature of $\theta_{\beta}$
by an appropriate rescaling $\theta_{\beta}=M\xi_{\beta}$, where
we choose (following \cite{Gromov:2006dh}) 
\[
M=-\frac{\log\mu}{2}\sim n_{0}\,.
\]
Then, according to \cite{Gromov:2006dh}, in the $n_{0}\rightarrow\infty$
limit the potential $V_{M}(\xi)=\frac{2\mu}{\pi}\cosh(M\xi)$ becomes
an infinitely deep (square) potential well confining $\xi$ to the
interval $(-2,2)$, and in this interval the equations become 
\begin{equation}
\frac{1}{M}\sum\limits _{\alpha}^{L}\frac{1}{\xi_{\beta}+\xi_{\alpha}}+\frac{1}{M}\sum\limits _{\alpha\neq\beta}^{L}\frac{1}{\xi_{\beta}-\xi_{\alpha}}+{\frac{1}{M\xi_{\beta}}\atop -\frac{1}{M\xi_{\beta}}}=-2m_{\beta}\,.\label{eqdupl}
\end{equation}
Note that in the limit $n_{0}\rightarrow\infty$, $n_{0}/M$ finite,
the last terms (originated in the interaction with the boundaries)
become sub-leading. Apart from the boundary contributions, this system
of equations is identical to the one considered in the periodic case
in \cite{Gromov:2006dh}, but for $2n_{0}$ particles located at $\xi_{\beta},\,-\xi_{\beta}$;
indeed changing $\xi_{\beta}\rightarrow-\xi_{\beta}$ in (\ref{eqdupl})
changes $m_{\beta}\rightarrow-m_{\beta}$, thus providing the (identical)
other half of the system. In \cite{Gromov:2006dh} the analogous system
was investigated by putting all mode numbers $m_{\beta}$ equal to
a common $m$, implying that in the continuous limit there is only
one $\xi$ cut. Looking at eq. (\ref{eqdupl}) as a system for $2n_{0}$
particles coming in pairs, shows that we cannot choose a common mode
number unless $m=0$. 

To implement the confining flat potential we introduce boundaries
with charges $q$ at $\xi=\pm2$, which we eventually will take to
zero. In the presence of the boundary charges $q$, the equilibrium
condition for the system of charges and mirror charges becomes 
\[
\sum\limits _{\alpha\neq\beta}^{n_{0}}\left(\frac{1}{\xi_{\beta}-\xi_{\alpha}}+\frac{1}{\xi_{\beta}+\xi_{\alpha}}\right)+{\frac{3}{2\xi_{\beta}}\atop -\frac{1}{2\xi_{\beta}}}=\frac{2qM\xi_{\beta}}{4-\xi_{\beta}^{2}}\qquad\beta=1,\dots,n_{0}\,.
\]
Now defining 
\[
Q(z)=z\prod\limits _{i}^{n_{0}}(z-\xi_{i})(z+\xi_{i})=z\prod\limits _{i}^{n_{0}}(z^{2}-\xi_{i}^{2}),\qquad\qquad P(z)=\frac{1}{z}\prod\limits _{i}^{n_{0}}(z^{2}-\xi_{i}^{2}),
\]
one readily proves, that 
\begin{eqnarray*}
\frac{Q^{\prime\prime}(\xi_{\beta})}{Q^{\prime}(\xi_{\beta})} & = & 2\left(\sum\limits _{\alpha\neq\beta}^{n_{0}}\left(\frac{1}{\xi_{\beta}-\xi_{\alpha}}+\frac{1}{\xi_{\beta}+\xi_{\alpha}}\right)+\frac{3}{2\xi_{\beta}}\right);\\
\frac{P^{\prime\prime}(\xi_{\beta})}{P^{\prime}(\xi_{\beta})} & = & 2\left(\sum\limits _{\alpha\neq\beta}^{n_{0}}\left(\frac{1}{\xi_{\beta}-\xi_{\alpha}}+\frac{1}{\xi_{\beta}+\xi_{\alpha}}\right)-\frac{1}{2\xi_{\beta}}\right).
\end{eqnarray*}
In the definition of $P,Q$ the product terms represent the $2n_{0}$
particles at $\pm\xi_{\beta}$, and the prefactors are introduced
to account for the boundary contributions. Though they look similar,
they are rather different: $Q(z)$ is a polynomial of order $2n_{0}+1$
having zeros at $z=0,\,\pm\xi_{i}$, while $P(z)$ is an analytical
function with zeros at $\pm\xi_{i}$, $z^{2n_{0}-1}$ asymptotic behaviour
at $z\rightarrow\infty$, and a single pole at $z=0$.

Let us analyze the free boundary condition with $m=0$. As a result
of the equilibrium conditions, the condition 
\[
(4-\xi_{\beta}^{2})Q^{\prime\prime}(\xi_{\beta})-4qM\xi_{\beta}Q^{\prime}(\xi_{\beta})=0
\]
holds (and similarly for $\xi_{\beta}\rightarrow-\xi_{\beta}$). Since
the polynomial $r(z)=(4-z^{2})Q^{\prime\prime}(z)-4qMzQ^{\prime}(z)$
has a zero at $z=0$, and is also of order $2n_{0}+1$, it must be
proportional to $Q(z)$. Matching the coefficients of the highest
powers of $z$ in them, we get 
\[
(4-z^{2})Q^{\prime\prime}(z)-4qMzQ^{\prime}(z)+(2n_{0}+1)(2n_{0}+4qM)Q(z)=0.
\]
With $z=2y$ this is the defining equation of the Jacobi polynomials
$P_{n}^{(\alpha,\beta)}(y)$ with $\alpha=\beta=2qM-1$, and $n=2n_{0}+1$.
Therefore 
\[
Q(z)=P_{2n_{0}+1}^{(2qM-1,2qM-1)}(\frac{z}{2}),
\]
and the $\xi_{i}$ are twice the positive roots of this polynomial.

Realizing that the ``free boundary'' resolvent $G(z)$ is related
to $Q(z)$ 
\begin{equation}
G(z)=\frac{1}{M}\left(\frac{1}{z}+\sum\limits _{i}^{L}\left(\frac{1}{z-\xi_{i}}+\frac{1}{z+\xi_{i}}\right)\right)=\frac{1}{M}\frac{Q^{\prime}(z)}{Q(z)},\label{freeG}
\end{equation}
one can derive an equation for it. Indeed using this relation and
the equilibrium conditions, we obtain 
\[
\frac{1}{M}G^{\prime}=-G^{2}+\frac{4qz}{4-z^{2}}G-\frac{(2n_{0}+1)(2n_{0}+4qM)}{M^{2}(4-z^{2})}.
\]
 In the continuum limit when $M\rightarrow\infty$, $n_{0}\rightarrow\infty$,
$n_{0}/M\sim{\cal O}(1)$, one can drop the l.h.s. and obtain an algebraic
equation for $G(z)$. Particularly interesting is the solution for
$q\rightarrow0$: 
\[
G(z)=\pm\frac{2n_{0}}{M}\frac{1}{z\sqrt{1-\frac{4}{z^{2}}}}.
\]
On the one hand it shows an inverse square root type singularity at
$z=\pm2$, on the other hand it differs only in the $n_{0}\rightarrow2n_{0}$
substitution from the analogous periodic expression \cite{Gromov:2006dh}.

Now we return to the investigation of the original problem eq. (\ref{eqdupl})
for the free boundary case. We choose a common mode number $m$ for
the positive solutions $\xi_{\beta}$ (likewise $-m$ for the negative
ones $-\xi_{\beta}$). Therefore in the continuum limit, the resolvent
has two cuts, one running between $(0,2)$, the other between $(-2,0)$,
i.e 
\[
G(z)=\frac{1}{2\pi}\int\limits _{0}^{2}dw\frac{\rho_{+}(w)}{z-w}+\frac{1}{2\pi}\int\limits _{-2}^{0}dw\frac{\rho_{-}(w)}{z-w}.
\]
$G(z)$, introduced explicitly in eq. (\ref{freeG}), satisfies $G(-z)=-G(z)$;
imposing this symmetry on this expression relates the two $\rho$-s
to each other 
\[
\rho_{-}(-w)=\rho_{+}(w)\equiv\rho(w),
\]
thus 
\[
G(z)=\frac{1}{2\pi}\int\limits _{0}^{2}dw\,\rho(w)\left(\frac{1}{z-w}+\frac{1}{z+w}\right)\,.
\]
In terms $G(z)$ one can write the continuum limit of eq.(\ref{eqdupl})
as 
\begin{equation}
\slash{\hspace{-0.2cm}}G(z)=G(z+i\epsilon)+G(z-i\epsilon)=-2m,\qquad x\in(0,2)\,.\label{Geq}
\end{equation}
We solve this by making an Ansatz for the $\rho(w)$ density motivated
by the finding of \cite{Gromov:2006dh} 
\[
\rho(w)=\frac{B}{\sqrt{4-w^{2}}}\,,
\]
where $B$ is a constant. Computing the principal value (PV) integral
one finds that $\slash{\hspace{-0.2cm}}G(z)=0,$ which becomes a constant
-- as required by eq.(\ref{Geq}) -- implying that $m=0$. $B$ is
determined from the $z\rightarrow\infty$ asymptotics of $G$: $G(z)\rightarrow\frac{2n_{0}+1}{M}\frac{1}{z}\sim\frac{2n_{0}}{M}\frac{1}{z}$,
leading to $B=\frac{4n_{0}}{M}$. With these values one finds (for
any $z$) 
\[
G(z)=\pm\frac{2n_{0}}{M}\frac{1}{z\sqrt{1-\frac{4}{z^{2}}}},
\]
which, after applying the Zhukowski map, $z=\lambda+\frac{1}{\lambda}$,
becomes 
\[
G(\lambda)=\pm\frac{2n_{0}}{M}\frac{\lambda}{\lambda^{2}-1}.
\]
We can see that $G(\lambda)$ is related to the (single) quasi-momentum
of the $O(4)$ $\sigma$-model with integrable (free) boundaries,
which corresponds to the uncharged, constant solution $n=0$. 

The situation with the mixed diagonal boundary, i.e. the one described
by the second version of eq. (\ref{eqdupl}) with the $-1/\xi_{\beta}$
boundary contribution, is a bit different from the free case. First,
we found numerical solutions for $m=0$ always with two imaginary
roots at $\xi_{0}=\pm i\alpha$. In the $M\to\infty$ and $n_{0}\to\infty$
limit, however, $\alpha\to0$ and we recover the free case. Defining
the resolvent as 
\[
\tilde{G}(z)=\frac{1}{M}\left(-\frac{1}{z}+\sum\limits _{i}^{n_{0}}\left(\frac{1}{z-\xi_{i}}+\frac{1}{z+\xi_{i}}\right)\right)\,,
\]
one can repeat the previous consideration leading to an identical
form for $\tilde{G}(z)$ in the continuum limit. The quasi-momenta
is related to the $n=0$ case in the classical analysis. To recover
the quasi-momenta of the one-cut solution, one has to introduce magnonic
roots and let them condense on an imaginary cut.

\section{Conclusions\label{sec:Conclusions}}

In this paper we analyzed the integrable boundary conditions of the
$O(N)$ non-linear $\sigma$-models at various levels.  Classically,
the models are conformal and conformality of the boundary condition implies the 
existence of a special set of infinitely many conserved charges.
Indeed, in the conformal boundary conditions, the boundary
limit of the difference of the light-cone components of the energy-momentum
tensor vanishes. In particular, this implies that the difference of
any integer power of the same light-cone components of the energy-momentum
tensor will also vanish, leading to an infinite family of conserved
charges. This conformality requirement can be guaranteed by connecting
the time derivative of the fundamental field to its space derivative
at the boundary by an anti-symmetric matrix, $M$ \cite{Moriconi:2001xz}.
By a similarity transformation, this matrix can be brought into a
2 by 2 block diagonal form, with different matrix entries. Taking
various limits of the matrix, one can recover boundary conditions
with Dirichlet and Neumann directions. In the coset formulation one
has to ensure that the time and space components of the conserved
currents are orthogonal for the trace at the boundary. This can be
achieved if one is obtained from the other by a commutation with another
matrix. However, this description is equivalent to the previous one
-- formulated on the fundamental fields -- only if the constraint
is added to the boundary Lagrangian. Thus the boundary condition in
\cite{Moriconi:2001xz} has to be modified with a non-linear term
(\ref{eq:BCMn}). 
Conformality of the boundary condition ensures the 
existence of infinitely many conserved charges. Whether these charges
are in involution or whether they provide enough conserved charges for the 
theory to be integrable is not investigated in the paper.

In order to classify classical solutions and to have a relation to
the quantum theory we introduced the boundary Lax formulation of the
problem. Integrable boundary conditions are classified by $O(N)$
group valued matrices, $U$, located at the boundaries. They are the
classical analogues of the reflection matrices and have to satisfy
an evolution equation (\ref{eq:Ui-restrictions-time-evolution}).
Unfortunately we could find only constant matrix solutions of these
equations, which are related to the mixture of Neumann and vanishing
Dirichlet boundary conditions. We characterized the analytical structure
and symmetry properties of the spectral curve of these boundary conditions. 

The quantization of the model is a highly non-trivial task. In the
bulk case the anomaly terms of the higher spin equations of motion
were classified and shown to have the same structure as the original
one, leading to quantum conservation laws. Unfortunately, anomalous
terms can appear at the boundary too, which can spoil the integrability
of all the boundary conditions. As we have no control over the boundary
anomalous terms, instead, assuming integrability, we classified the
quantum integrable boundary conditions by the solutions of the boundary
Yang-Baxter equations. We list all solutions found so far: these contain
diagonal reflection matrices with two different entries. They correspond
to Dirichlet and vanishing Neumann boundary conditions; there are
also boundary conditions with a single 2 by 2 block and otherwise
diagonal; in the $O(2N)$ case there is a boundary condition in which
all 2 by 2 blocks are the same. Bethe Ansatz equations in spin-chains
have been formulated in all cases, except when we have odd Dirichlet
(or Neumann) directions \cite{Arnaudon:2003gj}. We used them to formulate
the Bethe-Yang equations which determine the asymptotically large
volume spectrum of the models on the interval with identical boundary
conditions on the two sides. Since the classifications at the quantum
and classical levels do not match, we analyzed the classical limit
of certain solutions in the $O(4)$ models with all Neumann (free)
and two Dirichlet and two Neumann (mixed) boundary conditions. We
found that they correspond to the classical solution whose spectral
curve we previously calculated explicitly. 

There are many open questions. There is obviously a mismatch between
the boundary conditions found at the various levels.
This could be related to the fact that we analyzed 
in many cases conformal boundary conditions. It would be very interesting 
to see whether the corresponding special set of conserved charges do commute or, if we 
can find higher spin Casimir charges as well. 
It would be also challenging  to find time- and even field-dependent boundary $U$
matrices for any boundary conditions, which can be described by an
anti-symmetric matrix, $M$. The quantum theory suggests that they
may exist only for the cases of one single block or all blocks the
same. To get some insight one can calculate the classical limit of
all the Bethe-Yang equations including also magnonic roots and other
groups. In doing so, the derivation of the Bethe Ansatz equations
in the missing cases, i.e. when we have odd number of Neumann or Dirichlet
directions, are crucial.

It would be also of interest to figure out a quantisation of the boundary
system in which the anomalous terms can be directly calculated, and
the existence of quantum conservation laws can be explicitly checked.

\section*{Acknowledgments}

The work was supported by a Lendület and by the NKFIH 116505 Grant.
IA was supported by the NCN grant 2012/06/A/ST2/00396. We thank Gábor
Pusztai and Vidas Regelskis for the enlightening discussions. IA would
also like to thank the\emph{ Wigner Research Centre for Physics} as
well as the \emph{Institute for Theoretical Physics at Roland Eötvös
University }for hosting her during month-long collaboration visits.

\appendix

\section{Symmetries of the boundary conditions\label{sec:Symm-boundary-conds}}

In this appendix we investigate the residual symmetries of the model
(\ref{eq:Sxibdry}), for the various boundary conditions in the language
of the unconstrained variable. First we recall that as a consequence
of the (bulk) conservation of the currents, $\partial^{\alpha}J_{\alpha}^{A}=0$
($A=ij,\ iN$), the charges satisfy 
\[
\partial_{\tau}Q^{A}=\int\limits _{0}^{\pi}d\sigma\partial^{\tau}J_{\tau}^{A}=-\int\limits _{0}^{\pi}d\sigma\partial^{\sigma}J_{\sigma}^{A}=J_{\sigma}^{A}(\pi)-J_{\sigma}^{A}(0).\quad A=ij,\ iN\,.
\]
Assuming again the lack of interplay between the two boundaries (``locality''),
we conclude that those components of the bulk charges stay conserved
in the presence of boundaries for which $J_{\sigma}^{A}\vert=0$.

Now consider the case when $l$ ($0\leq l\leq N-1$) of the $\xi^{i}$-s
satisfy Neumann ($\partial_{\sigma}\xi^{i}=0$), while the rest are
Dirichlet ($\partial_{\tau}\xi^{i}=0$, $i=l+1,\dots,N-1$) b.c.-s.
To ease the notation we write the $\vxi$ field as $\vxi=(\vr,\vs)$
with boundary values $\partial_{\sigma}\vr=0$ and $\partial_{\tau}\vs=0$,
where $\vr=(r^{1},\dots,r^{l})$ and $\vs=(s^{1},\dots,s^{N-1-l})$.
(Note that in this notation $\xi^{2}=\vr^{2}+\vs^{2}$ and on the
boundary $\vxi\cdot\partial_{\sigma}\vxi=\vs\cdot\partial_{\sigma}\vs$,
$\vxi\cdot\partial_{\tau}\vxi=\vr\cdot\partial_{\tau}\vr$). This
notation is useful since the currents $J_{\alpha}^{ij}$ can be split
into three sets 
\[
J_{\alpha}^{ij}=\frac{4}{(1+\xi^{2})^{2}}(r^{i}\partial_{\alpha}r^{j}-r^{j}\partial_{\alpha}r^{i}),\ i,j=1,\dots,l;\quad J_{\alpha}^{l+i,l+j}=\frac{4}{(1+\xi^{2})^{2}}(s^{i}\partial_{\alpha}s^{j}-s^{j}\partial_{\alpha}s^{i})
\]
where $i,j=1,\dots,N-l-1$, and 
\[
J_{\alpha}^{i,l+j}=\frac{4}{(1+\xi^{2})^{2}}(r^{i}\partial_{\alpha}s^{j}-s^{j}\partial_{\alpha}r^{i}),\quad i=1,\dots,l,\ j=1,\dots,N-l-1,
\]
while those of $J_{\alpha}^{iN}$ split into two sets 
\[
J_{\alpha}^{iN}=-\frac{2}{(1+\xi^{2})^{2}}(2r^{i}\vxi\cdot\partial_{\alpha}\vxi+(1-\xi^{2})\partial_{\alpha}r^{i}),\quad i=1,\dots,l\,;
\]
\[
J_{\alpha}^{l+j,N}=-\frac{2}{(1+\xi^{2})^{2}}(2s^{j}\vxi\cdot\partial_{\alpha}\vxi+(1-\xi^{2})\partial_{\alpha}s^{j}).\quad j=1,\dots,N-l-1\,.
\]
It is obvious that the boundary values of $J_{\sigma}^{ij}\vert=0$
($ij=1,\dots,l$) as a result of the Neumann b.c. on $\vr$, and consequently
the $SO(l)$ symmetry generated by these currents survives. It is
slightly more complicated to see the symmetry coming from the fields
with Dirichlet b.c., however one can verify that the combinations
\[
\tilde{J}_{\sigma}^{l+i,l+j}=J_{\sigma}^{l+i,l+j}+\frac{2s^{i}}{1-\xi^{2}}\vert J_{\sigma}^{jN}-\frac{2s^{j}}{1-\xi^{2}}\vert J_{\sigma}^{iN}
\]
(where $\ \vert$ stand for the boundary values of the expressions
in question) do vanish on the boundary, and the $SO(N-1-l)$ symmetry
generated by them also survives. Thus the complete symmetry compatible
with the boundaries is $SO(l)\times SO(N-1-l)$ when $l$ of the $\xi^{i}$
fields satisfy Neumann while the remaining $N-1-l$ of them generic
Dirichlet b.c.-s.

If $l$ takes its maximal value $l=N-1$, then all $\vxi$ fields
satisfy Neumann b.c., and looking at (\ref{eq:nxirel}) we see that
also all components of $n^{I}$ do the same. From the expressions
of the currents above, it follows that all of the currents vanish
at the boundary, and the full $SO(N)$ symmetry of the bulk theory
is preserved by this b.c.. If, on the other hand, $l$ vanishes ($l=0$),
then all the $\vxi$ fields satisfy Dirichlet b.c., and from (\ref{eq:nxirel})
it follows that all components of $n^{I}$ do the same. From the previous
argument leading to $\tilde{J}_{\sigma}^{l+i,l+j}$ above, it follows
that the symmetry of this ``all Dirichlet'' b.c. is $SO(N-1)$,
i.e. the bulk $SO(N)$ symmetry is broken.

In the following we derive the boundary conditions for the constrained
variables. Without loss of generality we assume that $\vs\vert=(\alpha,0,\dots,0)$.
This, in particular, implies the vanishing Dirichlet boundary condition
for $n_{i}$: 
\[
n_{i}\vert=\frac{2s_{i}}{1+r^{2}+\alpha^{2}}\vert=0\quad;\qquad i=l+2,\dots,N-1
\]
Let us introduce primed variables by the combinations 
\begin{eqnarray*}
n_{l+1}' & = & \frac{1}{\sqrt{1+\alpha^{2}}}\left(n_{l+1}-\alpha n_{N}\right);\\
n_{N}' & = & \frac{1}{\sqrt{1+\alpha^{2}}}\left(\alpha n_{l+1}+n_{N}\right),
\end{eqnarray*}
as well as $n_{i}'=n_{i}$ for $i\neq l+1,\, N$. One can easily see
that $n_{l+1}'$ satisfies Dirichlet boundary condition 
\[
n_{l+1}'\vert=\frac{\alpha}{\sqrt{1+\alpha^{2}}}.
\]
Actually the primed coordinates have length $1$ and can be obtained
from $n_{i}$ by an ortogonal transformation. One can check that from
these new coordinates, $N-l-1$ satisfy a Dirichlet boundary condition,
while $l+1$ satisfy the generalized Neumann boundary condition (\ref{eq:gen_neum_bc}).
This makes the two formulations completely equivalent. Choosing, in
particular, $\forall s^{j}\vert=0$ (i.e. $\vs\vert=0$, ``vanishing''
Dirichlet) makes both $\partial_{\sigma}n^{i}\vert=0$ and $\partial_{\tau}n^{l+j}\vert=0$
in addition to guaranteeing $\partial_{\sigma}n^{N}\vert=0$. Looking
at the previous expressions for the various currents reveals that
for this vanishing Dirichlet b.c. not only $J_{\sigma}^{l+i,l+j}$
vanish on the boundary but also $J_{\sigma}^{iN}$ (while $J_{\sigma}^{l+j,N}\vert\neq0$).
The -- now conserved -- $J_{\alpha}^{iN}$ combine with $J_{\alpha}^{ij}$
to generate an $SO(l+1)$ symmetry. Thus in this case the total symmetry
compatible with the boundaries is $SO(l+1)\times SO(N-1-l)$.

Now that we determined the symmetries compatible with the consistent
boundary conditions and the (matrix) form of $J_{\sigma}$ and $J_{\tau}$
on the boundary, we can search for a constant $U$ matrix which satisfies
(\ref{eq:Ui-commutation-Js}). When the remaining symmetry is $SO(l+1)\times SO(N-1-l)$,
the construction of such a $U$ is known from the mathematical literature
\cite{MR1834454}, since $SO(N)/SO(l+1)\times SO(N-1-l)$ is a symmetric
space. However, when the residual symmetry is $SO(l)\times SO(N-1-l)$,
no such $U$ matrix exists, since $SO(N)/SO(l)\times SO(N-1-l)$ is
NOT a symmetric space for $1\leq l<N-1$. This would be an example
when a model is conformal but  not Lax
integrable.

The b.c.-s (\ref{eq:teljes}) for the $\vxi$ fields can be translated
into b.c.-s for the currents $J_{\sigma}^{ij}$ and $J_{\sigma}^{iN}$:
we take eq. (\ref{eq:aram1}) and (\ref{eq:aram2}) and replace $\partial_{\sigma}\xi^{k}\vert$
in them by the r.h.s. of eq. (\ref{eq:teljes}); then, after some
algebraic manipulations, we try to identify $J_{\tau}^{ij}$ and $J_{\tau}^{iN}$
in what is obtained. Since we previously derived the b.c. 
\begin{equation}
J_{\sigma}^{IJ}=\frac{1}{2}\Bigl([M,J_{\tau}]+[mMm,J_{\tau}]\Bigr)^{IJ},\quad m^{IJ}=\delta^{IJ}-2n^{I}n^{J}\,,\label{eq:tamasbc}
\end{equation}
we need only to check whether this condition is a consequence of (\ref{eq:teljes}).
A direct computation gives 
\begin{eqnarray*}
\Bigl([M,J_{\tau}]+[mMm,J_{\tau}]\Bigr)^{im} & = & \frac{8}{(1+\xi^{2})^{2}}\Bigl(\xi^{i}M_{ml}\dot{\xi}^{l}-\xi_{m}M_{il}\dot{\xi}^{l}\\
 &  & \,\,\,\,\,\,\,\,\,\,\,\,\,\,\,\,\,\,\,\,\,\,\,\,\,\,\,\,\,-\frac{2\vxi\cdot\dot{\vxi}}{1+\xi^{2}}\bigl(\xi^{i}(M_{ml}\xi^{l}+M_{mN})-\xi^{m}(M_{il}\xi^{l}+M_{iN})\bigr)\Bigr)\,,
\end{eqnarray*}
\begin{eqnarray*}
\Bigl([M,J_{\tau}]+[mMm,J_{\tau}]\Bigr)^{iN} & = & \frac{4}{(1+\xi^{2})^{2}}\Bigl(2\xi^{i}M_{Nl}\dot{\xi}^{l}-(1-\xi^{2})M_{il}\dot{\xi}^{l}\\
 &  & \,\,\,\,\,\,\,\,\,\,\,\,\,\,\,\,\,\,\,\,\,\,\,\,\,\,\,\,\,+\frac{2\vxi\cdot\dot{\vxi}}{1+\xi^{2}}\bigl((1-\xi^{2})(M_{il}\xi^{l}+M_{iN})+2\xi^{i}M_{Nl}\xi^{l}\bigr)\Bigr)\,,
\end{eqnarray*}
and after a not very informative computation one obtains that our
currents do indeed satisfy (\ref{eq:tamasbc}).

Next we translate these b.c.-s for boundary conditions for the $n^{I}$
fields. Starting from eq. (\ref{eq:teljes}) one finds 
\[
\partial_{\sigma}n^{K}=\bigl(\sum\limits _{I}M_{IK}\dot{n}^{I}-n^{K}\sum\limits _{L,I}n^{L}M_{IL}\dot{n}^{I}\bigr)\,.
\]
This b.c. is highly non-linear and depends also on the boundary values
of $n^{I}$, not only on its derivatives. The first term on the r.h.s.
gives the naïve (and inconsistent) b.c. derived by Moriconi \cite{Moriconi:2001xz}
and analyzed by He and Zhao \cite{He:2003iw}, and the second non-linear
term guarantees that this b.c. is consistent, i.e. the r.h.s vanishes
also when it is multiplied by $n^{K}$ and summed over $K$. For $N=3$,
when the antisymmetric $M_{IJ}$ can be described in terms of a three
vector $\vec{q}=(q^{1},q^{2},q^{3})$ as $M_{IJ}=\epsilon_{IJK}q^{K}$,
this b.c. is identical to the one derived by Corrigan and Sheng \cite{Corrigan:1996nt}.

To see what residual symmetries are compatible with eqs. (\ref{eq:teljes})
and (\ref{eq:tamasbc}) we determine which parts of the bulk $SO(N)$
transformations leave the boundary Lagrangian $\sum_{IJ}n^{I}M_{IJ}\dot{n}^{J}$
invariant. Since the infinitesimal $SO(N)$ transformations can be
written as 
\[
n^{K}\rightarrow n^{K}+\epsilon^{AB}\Sigma_{AB}^{KL}n^{L},\qquad\Sigma_{AB}^{KL}=\delta_{A}^{K}\delta_{B}^{L}-\delta_{A}^{L}\delta_{B}^{K},\qquad\epsilon^{AB}=-\epsilon^{BA},
\]
one readily obtains that the transformations commuting with $M$,
\[
[M,\epsilon^{AB}\Sigma_{AB}]=0,
\]
are the ones that leave the boundary Lagrangian invariant. Since,
by an appropriate orthogonal transformation, any antisymmetric $N\times N$
matrix can be brought to the block diagonal form 
\[
M_{IJ}=\left(\begin{array}{ccccccc}
0 & m_{1}\\
-m_{1} & 0\\
 &  & 0 & m_{2}\\
 &  & -m_{2} & 0\\
\\
\\
 &  &  &  &  &  & 0
\end{array}\right)
\]
(all $m_{i}$-s, $i=1,\dots\Bigl[\frac{N}{2}\Bigr]$, are real) it
is enough to analyze the symmetries of the model when we use these
matrices in the boundary Lagrangian. If we assume that the first $k$
$m_{i}$-s are all different $m_{i}\neq m_{j}$ $i,j=1,\dots k$ and
non zero, while the rest vanishes, then the subgroup of $SO(N)$ commuting
with this $M$ is $\underbrace{SO(2)\times\dots SO(2)}_{k}\times SO(N-2k)$.
If, on the other hand, $N=2L$ is even, and all $m_{i}$-s are equal
and non-vanishing, then the subgroup of $SO(2L)$ commuting with this
$M$ is $U(L)$ \cite{MR1834454}.

\section{Comments on boundary conditions with not-maximal radius}

In this appendix we consider the boundary condition when the fields
are restricted to a sphere $S^{n-k-1}$ of radius $\cos{\phi}$. In
this case the constraints at the boundary are the following: 
\begin{align}
\nt^{t}\nt & =\cos(\phi)^{2},\label{curve}\\
\nh^{t}\nh & =\sin(\phi)^{2},\nonumber \\
\partial_{0}\nh & =0.\nonumber 
\end{align}
 If we want to use stereographic projection to obtain the unconstrained
variables in the bulk we have to introduce some extra constraints
at the boundary, as (\ref{curve}) is not a flat hypersurface in $\mathbb{R}^{n-1}$.
We can use the following coordinates: 
\begin{align*}
\nh & =(0,\dots,0,\sin(\phi)),\\
n_{i} & =\frac{2\xi_{i}}{1+\xi^{2}},\quad n_{n}=\frac{1-\xi^{2}}{1+\xi^{2}},\quad\xi^{2}=\xi_{i}\xi_{i}.
\end{align*}
 At the boundary, however, the $\xi_{i}$-s live only on an $(n-k-1)$-dimensional
sphere because 
\[
\xi^{2}=\frac{1-\sin(\phi)}{1+\sin(\phi)}.
\]
We can see that the $\xi_{i}$-s are subject to a constraint at the
boundary. Therefore, in the language of $\xi_{i}$-s and $\n$-s this
boundary condition cannot be written in any simple homogeneous Neumann
and Dirichlet form.

In the coset language there is a subalgebra $\hn$ (algebra of the
little group) where $\hn\n_{0}=0$ and another subalgebra $\gn_{1}$
(which generates the $S^{n-k-1}$, $G_{1}=\On{n-k}$) where $\nt=\exp(\gn_{1})\n_{0}$.
In this case the reference vector is $\n_{0}=(\cos(\phi),0,\dots,0,\sin(\phi))$.

We can decompose $\gn_{1}$ to $\hn_{1}\oplus\fn_{1}$, where $\hn_{1}\subset\hn$,
and we can convince ourselves that $\fn_{1}$ is not a subset of $\fn$.
Therefore the algebra of the currents cannot be decomposed into a
$\mathbb{Z}_{2}\times\mathbb{Z}_{2}$ graded algebra, which means
that the boundary conditions cannot be written as a commutator and
an anticommutator.

\end{document}